\documentclass[journal,twoside,web]{ieeecolor}
\usepackage{generic}
\usepackage{cite}
\usepackage{amsmath,amssymb,amsfonts,mathrsfs}
\usepackage{algorithm}
\usepackage{graphicx}
\let\labelindent\relax
\usepackage{textcomp,enumitem}
\usepackage{float}
\usepackage{subfig}
\usepackage[justification=centering]{caption}
\usepackage{bm}
\usepackage{algpseudocode}
\usepackage{color}
\usepackage[dvipsnames]{xcolor}
\newtheorem{thm0}{Theorem}
\newtheorem{Theorem 1}[thm0]{Theorem}
\newtheorem{asm}{Assumption}
\newtheorem{Assumption 1}[asm]{Assumption}
\newtheorem{def0}{Definition}
\newtheorem{rmk}{Remark}
\newtheorem{Remark 1}[rmk]{Remark}
\newtheorem{lemma}{Lemma}
\newtheorem{Lemma 1}[lemma]{Lemma}
\newtheorem{Cor}{Corollary}
\newtheorem{Corollary 1}[Cor]{Corollary}

\newcommand{\diag}{\textrm{diag}}
\newcommand{\p}{\mathbb{P}}
\newcommand{\E}{\mathbb{E}}
\def\BibTeX{{\rm B\kern-.05em{\sc i\kern-.025em b}\kern-.08em
    T\kern-.1667em\lower.7ex\hbox{E}\kern-.125emX}}
\begin{document}
\title{\bf Differentially Private Distributed Nonconvex Stochastic Optimization with Quantized Communication \thanks{The work was supported by National Natural Science Foundation of China under Grants 62203045, 62433020 and T2293770. The material in this paper was not presented at any conference.}}
\author{Jialong Chen, Jimin Wang, \IEEEmembership{Member, IEEE}, and Ji-Feng Zhang, \IEEEmembership{Fellow, IEEE}
\thanks{Jialong Chen is with the State Key Laboratory of Mathematical Sciences, Academy of Mathematics and Systems Science, Chinese Academy of Sciences, Beijing 100190, and also with the School of Mathematical Sciences, University of Chinese Academy of Sciences, Beijing 100049, China. (e-mail: chenjialong23@mails.ucas.ac.cn)}
\thanks{Jimin Wang is with the School of Automation and Electrical Engineering, University of Science and Technology Beijing, Beijing 100083, and also with the Key Laboratory of Knowledge Automation for Industrial Processes, Ministry of Education, Beijing 100083, China (e-mail: jimwang@ustb.edu.cn)}
\thanks{Ji-Feng Zhang is with the School of Automation and Electrical Engineering, Zhongyuan University of Technology, Zheng Zhou 450007; and also with the State Key Laboratory of Mathematical Sciences, Academy of Mathematics and Systems Science, Chinese Academy of Sciences, Beijing 100190, China. (e-mail: jif@iss.ac.cn)}}

\maketitle
\begin{abstract}
	This paper proposes a novel distributed nonconvex stochastic optimization algorithm that can achieve privacy protection and convergence simultaneously while improving communication efficiency. Specifically, each node adds general privacy noises to its local state to avoid information leakage, and then, quantizes its noise-perturbed state before transmitting to improve communication efficiency. By using a sampling parameter-controlled subsampling method, the proposed algorithm enhances the differential privacy level compared to the existing works. By using a new convergence analysis technique, the mean square convergence for nonconvex cost functions is given without assuming that gradients are bounded. Furthermore, when the nonconvex cost function satisfies the Polyak-{\L}ojasiewicz condition, a convergence rate and the oracle complexity of the proposed algorithm are given. By using a two-time-scale step-sizes method and a probabilistic quantizer, the proposed algorithm achieves finite cumulative differential privacy budgets $\epsilon$, $\delta$ and the mean square convergence simultaneously while improving communication efficiency as the sample-size goes to infinity. A numerical example of the distributed training on the ``MNIST'' dataset is given to show the effectiveness and advantages of the algorithm.
\end{abstract}

\vspace{-1em}
\IEEEpeerreviewmaketitle
\section{Introduction}
\IEEEPARstart{D}{istributed} optimization is gaining more and more attraction due to its fundamental role in cooperative control, smart grids, sensor networks, and large-scale machine learning (\!\!\cite{zhu2011distributed,zhu2012approximate,doan2021a,doan2021b,xin2022fast,jiang2017collaborative,lu2023convergence,zhang2023communication,ge2023accelerate,lei2022distributed}).  As an important type of distributed optimization,  distributed stochastic optimization has gained popularity due to its superior performance in handling stochastic cost functions (\!\cite{lu2023convergence,jiang2017collaborative,zhang2023communication,ge2023accelerate,lei2022distributed}). So far, substantial efforts have been dedicated to the field of distributed stochastic optimization for both convex cost functions (e.g., \cite{jiang2017collaborative,lei2022distributed}) and nonconvex cost functions (e.g., \cite{lu2023convergence,zhang2023communication,ge2023accelerate}). In practice, nonconvex cost functions has wider applications than convex cost functions. For example, cost functions are often nonconvex in the training of recurrent neural networks (\!\!\cite{pascanu2013difficulty}) and the policy optimization of linear quadratic regulator (\!\!\cite{fazel18global}). It is worth mentioning that saddle points in nonconvex cost functions may cause sharp changes of gradients (\!\!\cite{pascanu2013difficulty,ge2015escaping}), and thus, pose the difficulty in the convergence (\!\!\cite{ge2023accelerate,zhang2023communication,jiang2017collaborative}).  To prove the convergence, it usually requires the assumption that the gradients are bounded (\!\!~\cite{lu2023convergence}), which is hard to be satisfied or verified in many practical scenarios (\!\!\cite{pascanu2013difficulty,fazel18global}). 

When studying distributed stochastic optimization problems, there are two key issues worthy of attention. One is the network bandwidth limitation, and the other is the leakage of the sensitive information concerning cost functions.  To solve the first issue, a common method is to transmit quantized information instead of the raw information. Generally, as a data compression technique to conserve network bandwidth, plenty of quantizers have been successfully applied to distributed stochastic optimization,  such as  the probabilistic quantizer (\!\!~\cite{zhang2023communication}),  the cluster-aware sketch based quantizer (\!\!~\cite{ge2023accelerate}), the uniform quantizer (\!\!\cite{rabbat2005quantized}), the Lloyd-Max quantizer (\!\!\cite{chen2024communication}).  These works have made significant contributions to improving communication efficiency,  neglecting the guarantee of convergence.  Taking \cite{rabbat2005quantized,chen2024communication} as examples, the amount of energy and bandwidth used for communication have been greatly reduced. However, the convergence cannot  be  guaranteed due to the biased quantization error (\!\!\cite{rabbat2005quantized}), and the unbounded variance of the quantization error (\!\!\!\cite{chen2024communication}).  It is noteworthy that eliminating the effect of the quantization error on the convergence for distributed stochastic optimization is nontrivial.  Fortunately, by a novel adaptive level quantizer, the convergence is achieved with requiring the increasing network bandwidth in \cite{faghri2020adaptive}.  This requirement restricts the applicability of this method. Recently,  by using the probabilistic quantizer, \cite{reisizadeh2019robust}  achieves  the convergence and communication efficiency simultaneously. 

To solve the second issue, it needs to design privacy-preserving techniques to protect the sensitive information (\!\!~\cite{zhang2021}). So far, various techniques have been employed to protect the sensitive information,  such as homomorphic encryption (\!\!\cite{lu2018privacy}), adding a constant uncertain parameter in step-sizes (\!\!\cite{lou2018privacy}), state decomposition (\!\!\cite{wang2019privacy}), adding deterministic perturbations in input and output (\!\!\cite{lu2020privacy}), adding noises (\!\!\cite{mo2017privacy,le2013differentially,dwork2014algorithmic,liu2020differentially,wang2022differentially1,chen2023locally,wang2022differentially2}) and so on.  Due to its simplicity in realization and immunity to post-processing, differential privacy has attracted a lot of attention and been used to solve privacy issues in distributed stochastic optimization (\!\cite{zhang2018improving,li2018differentially,huang2019dp,ding2021differentially,gratton2021privacy,xu2022dp,yan2023killing,wang2023quantization,wang2023decentralized,liu2024distributed}). One commonly used differential privacy in distributed stochastic optimization is $(\epsilon,\delta)$-differential privacy achieved by the Laplacian or (discrete) Gaussian noise (\!\cite{zhang2018improving,li2018differentially,ding2021differentially,huang2019dp,gratton2021privacy,liu2024distributed,xu2022dp,wang2023decentralized}), the binomial-mechanism-aided quantizer (\!~\cite{yan2023killing}).  Among others, \cite{xu2022dp,gratton2021privacy, yan2023killing} achieve  $(\epsilon,\delta)$-differential privacy while sacrificing the convergence accuracy, which is undesirable in accuracy-sensitive applications.  To tackle this  dilemma,  some novel methods have been proposed in \cite{zhang2018improving,liu2024distributed,li2018differentially,huang2019dp,ding2021differentially,wang2023decentralized} under the assumption that gradients are bounded.  For instance,  by introducing a weakening factor to mitigate the impact of  decaying privacy noises (\!\!~\cite{zhang2018improving,huang2019dp}) or constant  privacy noises (\!\!\cite{li2018differentially,liu2024distributed,wang2023decentralized}),  the convergence and $(\epsilon,\delta)$-differential privacy are achieved simultaneously.  By proposing an iteration maximum-based method, the convergence is achieved with enhanced differential privacy in \cite{ding2021differentially}. Although the analysis is elegant in  \cite{zhang2018improving,li2018differentially,huang2019dp,ding2021differentially,gratton2021privacy,xu2022dp,wang2023decentralized,liu2024distributed,yan2023killing}, differential privacy budgets  go to infinity over infinite iterations, and thus, the sensitive information therein cannot be protected over infinite iterations.  
By making interesting connections to the stochastic quantizer,  $(0,\delta)$-differential privacy is proved in distributed stochastic optimization.  A pioneering work in this direction is \cite{wang2023quantization},  where $(0,\delta)$-differential privacy is achieved by the ternary quantizer at each iteration. This implies that $(0,1)$-differential privacy is achieved over infinite iterations. Since $(0,1)$-differential privacy means the algorithm directly outputs the sensitive information, the sensitive information therein cannot be protected over infinite iterations.  

Although some advancements have been made for considering network bandwidth limitation and privacy preserving simultaneously for distributed stochastic optimization (\!\!\cite{ding2021differentially,wang2023quantization,yan2023killing}), some open problems still persist.  \cite{yan2023killing} provides new insights into the correlated nature of communication and privacy,  but the convergence accuracy is sacrificed.  \cite{ding2021differentially,wang2023quantization} proposes a comprehensive solution that could simultaneously achieve privacy preserving, convergence and improved communication efficiency under the assumption that gradients are bounded.  Regarding the privacy preserving, the sensitive information in \cite{ding2021differentially,wang2023quantization,yan2023killing} cannot be protected over infinite iterations.
 
Motivated by the aforementioned observations, the following questions may be raised:
``how to design a privacy-preserving distributed nonconvex stochastic optimization algorithm that can enhance the differential privacy level while achieving convergence and improving communication efficiency simultaneously, especially avoiding the bounded gradients often occurred in existing results?" If there exists such an algorithm, then we are further concerned about ``how do the added privacy noises affect the convergence rate of the algorithm?" In this paper, we give analytical solutions to the above questions and propose a novel differentially private distributed nonconvex stochastic optimization algorithm with quantized communication. The main contribution is as follows:\vspace{-0.2em}
\begin{itemize}[leftmargin=*]
\item A sampling parameter-controlled subsampling method is proposed to enhance the differential privacy level. By using this subsampling method, cumulative differential privacy budgets $\epsilon$, $\delta$ are reduced with guaranteed mean square convergence for general privacy noises.  Furthermore,  finite cumulative differential privacy budgets $\epsilon$, $\delta$ are achieved over infinite iterations. 

\item In comparison to the existing results, the mean square convergence of the algorithm for nonconvex cost functions is achieved by removing the assumption that gradients are bounded. Furthermore, when the nonconvex cost function satisfies the Polyak-{\L}ojasiewicz condition, a convergence rate of the algorithm for general privacy noises is provided, including decaying, constant and increasing privacy noises. This is non-trivial even without considering privacy protection problem. 

\item A two-time-scale step-sizes method is employed to eliminate the effect of the quantization error on the convergence. By combining this method with a probabilistic quantizer, the mean square convergence of the algorithm is achieved while improving communication efficiency simultaneously.  More interestingly, finite cumulative differential privacy budgets $\epsilon$, $\delta$ over infinite iterations and the mean square convergence of the algorithm are achieved simultaneously while improving communication efficiency for the first time.

\item The effectiveness of our algorithm is evaluated by using the distributed training of a convolutional neural network on the ``MNIST'' dataset. Our experimental results confirm that the proposed approach is superior to existing counterparts in terms of training/test accuracies,  convergence rate, and differential privacy level.
\end{itemize}\vspace{-0.2em}

%The results in this paper are significantly different from those in existing works. A comparison with the state-of-the-art is as follows: Compared with \cite{zhang2018improving,li2018differentially,huang2019dp,ding2021differentially,gratton2021privacy,xu2022dp,wang2023quantization,wang2023decentralized,yan2023killing,liu2024distributed}, finite cumulative differential privacy budgets $\epsilon$, $\delta$ are achieved over infinite iterations. Compared with \cite{lu2023convergence,ge2023accelerate,zhang2018improving,li2018differentially,huang2019dp,ding2021differentially,gratton2021privacy,wang2023quantization,wang2023decentralized,liu2024distributed}, the convergence of the proposed algorithm is given. Compared with \cite{ge2023accelerate,gratton2021privacy,zhang2023communication,yan2023killing}, the convergence is achieved while achieving differential privacy. Compared with  \cite{jiang2017collaborative,zhang2023communication,ge2023accelerate,lu2023convergence,zhang2018improving,li2018differentially,huang2019dp,gratton2021privacy,lei2022distributed,xu2022dp,wang2023decentralized,liu2024distributed}, privacy protection and  are considered simultaneously in this paper.

This paper is organized as follows: Section \ref{section2} formulates the problem to be investigated. Section \ref{section3} presents the main results including the privacy, convergence and oracle complexity analysis of the algorithm. Section \ref{section4} provides a numerical example. Section \ref{section5} gives some concluding remarks.

\emph{Notation:}
$\mathbb{R}$ and $\mathbb{R}^{r}$ denote the set of all real numbers and $r$-dimensional Euclidean space, respectively. $\text{Range}(F)$ denotes the range of a mapping $F$, and $F\circ G$ denotes the composition of mappings $F$ and $G$. For sequences $\{a_k\}_{k=1}^{\infty}$ and $\{b_k\}_{k=1}^{\infty}$, $a_k=O(b_k)$ means there exists $A_1\ge0$ such that $\limsup_{k\to\infty}|\frac{a_k}{b_k}|\leq A_1$. $\mathbf{1}_n$ represents an $n$-dimensional vector whose elements are all 1. $A^\top$ stands for the transpose of the matrix $A$. We use the symbol $\|x\|=\sqrt{x^{\top}x}$ to denote the standard Euclidean norm of $x=[x_1,x_2,\dots,x_m]^\top$, and $\|A\|$ to denote the 2-norm of the matrix $A$. $\mathbb{P}(\mathcal{B})$ and $\mathbb{E}(X)$ refer to the probability of an event $\mathcal{B}$ and the expectation of a random variable $X$, respectively. $\otimes$ denotes the Kronecker product of matrices. $\lfloor z \rfloor$ denotes the largest integer no larger than $z$. For a vector $v=[v_1,v_2,\dots,v_n]^\top$, $\diag(v)$ denotes the diagonal matrix with diagonal elements being $v_1,v_2,\dots,v_n$. For a differentiable function $f(x)$, $\nabla f(x)$ denotes its gradient at the point $x$.\vspace{-0.8em}
\section{Preliminaries and problem formulation}\label{section2}
\subsection{Graph theory}
Consider a network of $n$ nodes which exchange information on an undirected and connected communication graph $\mathcal{G}=(\mathcal{V},\mathcal{E})$. $\mathcal{V}=\{1,2,\dots,n\}$ is the set of all nodes, and $\mathcal{E}$ is the set of all edges. An edge $e_{ij}\in \mathcal{E}$ if and only if Node $i$ can receive the information from $j$. Different nodes in $\mathcal{V}$ exchange information based on the weight matrix $\mathcal{A}=\left(a_{ij}\right)_{1\leq i,j\leq n}$, whose entry $a_{ij}$ is either positive if $e_{ij}\in \mathcal{E}$, or 0, otherwise. The neighbor set of Node $i$ is defined as $\mathcal{N}_{i}$$=$$\{j$$\in$$\mathcal{V}$$:a_{ij}>0\}$, and the Laplacian matrix of $\mathcal{A}$ is defined as $\mathcal{L}=\diag(\mathcal{A}\mathbf{1}_n)-\mathcal{A}$. The assumption about the weight matrix $\mathcal{A}$ is given as follows:
\begin{asm}\label{assumption 1: graph preliminaries}
	The weight matrix $\mathcal{A}$ is doubly stochastic, i.e., $\mathcal{A}\mathbf{1}_n=\mathbf{1}_n$, $\mathbf{1}_n^\top\mathcal{A}=\mathbf{1}_n^\top$.
\end{asm}
\begin{rmk}
	Assumption \ref{assumption 1: graph preliminaries} is standard and commonly used in undirected and connected communication graphs (see e.g. \cite{doan2021a,doan2021b,jiang2017collaborative,mo2017privacy,ding2021differentially,huang2019dp,gratton2021privacy,xu2022dp,wang2023decentralized,wang2023quantization}). There are many examples satisfying Assumption \ref{assumption 1: graph preliminaries} in practice, such as, the dynamic load balancing of distributed memory \mbox{processors \cite{cybenko1989dynamic}}, the distributed estimation of sensor networks \cite{blatt2004distributed} and the distributed machine learning \cite{lian2017can}.
\end{rmk}\vspace{-1em}
\subsection{Distributed stochastic optimization}
In this paper, the following distributed nonconvex stochastic optimization problem is considered:
\begin{align}\label{question of interest}
	\smash{\min_{x\in\mathbb{R}^{r}}\!F(x)\!=\!\min_{x\in\mathbb{R}^{r}}\!\frac{1}{n}\!\sum_{i=1}^{n}\!f_{i}(x), f_{i}(x)\!=\!\mathbb{E}_{\xi_{i}\sim \mathscr{D}_{i}}[\ell_{i}(x,\xi_{i})],}
\end{align}
{\vskip 2pt}\noindent where $x$ is available to all nodes, $\ell_{i}(x,\xi_{i})$ is a local cost function which is private to Node $i$, and $\xi_i$ is a random variable drawn from an unknown probability distribution $\mathscr{D}_i$. In practice, since the probability distribution $\mathscr{D}_i$ is difficult to obtain, it is replaced by the dataset $\mathcal{D}_i$$=$$\{$$\xi_{i,l}$$\in$$\mathbb{R}^p$$:l=1,\dots, D\}$. Then, \eqref{question of interest} is rewritten as the following empirical risk minimization problem:
\begin{align}\label{question of interest 2}
		\smash{\min_{x\in\mathbb{R}^{r}}\!F(x)\!=\!\min_{x\in\mathbb{R}^{r}}\!\frac{1}{n}\!\sum_{i=1}^{n}\!f_{i}(x), f_{i}(x)\!=\!\frac{1}{D}\sum_{l=1}^{D}\ell_{i}(x,\xi_{i,l}).}
\end{align}
{\vskip 2pt}\noindent When solving the empirical risk minimization problem \eqref{question of interest 2}, a stochastic first-order oracle is often required (\!\!\cite{bubeck2015convex}), which returns a sampled gradient $\nabla\ell_i(x,\zeta_i)$ of $f_{i}(x)$ for any $i$$\in$$\mathcal{V}$, $x$$\in$$\mathbb{R}^{r}$ and $\zeta_{i}$ uniformly sampled from $\mathcal{D}_i$. Then, the following standard assumption is given:
\begin{asm}\label{assumption 2: cost functions}
	(i) There exists $L_1,L_2>0$ such that for any $i\in\mathcal{V}$, $\ell_i(x,\zeta_i)$ is $L_1$- and $L_2$-smooth with respect to $x$ and $\zeta_i$, respectively, i.e., $\|\nabla \ell_i(x,\zeta_i)\!-\!\nabla \ell_i(y,\zeta_i)\|\!\leq\! L_1\|x\!-\!y\|$, $\|\nabla \ell_i(x,\zeta_i)\!-\!\nabla \ell_i(x,\zeta^\prime_i)\|\!\leq\! L_2\|\zeta_i\!-\!\zeta^\prime_i\|$, $\forall x,y$$\in$$\mathbb{R}^{r}$,  $\forall \zeta_i,\zeta^\prime_i$$\in$$\mathbb{R}^p$.\vspace{-0.2em}
	\begin{enumerate}[leftmargin=*,label=(\roman{*})]
		\setcounter{enumi}{1}
		\item Each cost function is bounded from below, i.e., $\min_{x\in\mathbb{R}^{r}}\!f_i(x)\!=\!f_i^*\!>\!-\infty$.
		\item There exists $\sigma_{\ell}$$>$$0$ such that each sampled gradient $\nabla\ell_i(x,\zeta_{i})$ satisfies $\mathbb{E}[\nabla\ell_i(x,\zeta_{i})]=\nabla f_i(x)$, $\mathbb{E}$$[\|\nabla\ell_i(x,\zeta_{i})-\nabla f_i(x)\|^2]$$\leq$$\sigma_{\ell}^2$.
	\end{enumerate}
\end{asm}
\begin{rmk}\label{rmk2}
	Assumption \ref{assumption 2: cost functions}(i) is commonly used (see e.g., \cite{zhang2023communication,pascanu2013difficulty,ge2015escaping,zhang2018improving,huang2019dp,wang2023decentralized}). Assumption \ref{assumption 2: cost functions}(ii) ensures the existence of the optimal solution. Assumption \ref{assumption 2: cost functions}(iii) requires that each sampled gradient $\nabla\ell_i(x,\zeta_{i})$ is unbiased with a bounded variance $\sigma_{\ell}^2$ (see e.g.\mbox{\cite{lei2022distributed,ding2021differentially,xu2022dp,wang2023quantization,yan2023killing}}).
\end{rmk}\vspace{-1em}
\subsection{Quantized communication}
Due to the network bandwidth limitation, the exchange of the uncompressed information brings communication burden. To address this, the probabilistic quantizer is used to quantize the exchanged information in this paper, which is a randomized mapping that maps an input to different values in a discrete set with some probability distribution, and satisfies the following assumption:
\begin{asm}\label{assumption 3: quantizer}
		The probabilistic quantizer $Q(x)$ is unbiased and its variance is bounded, which means there exists $\Delta>0$, such that $\E(Q(x)|x)=x$ and $\E(|Q(x)\!-\!x|^2|x)\!\le\!\Delta^2$.
\end{asm}
\begin{rmk}
		Assumption \ref{assumption 3: quantizer} is standard and commonly used (see e.g. \cite{ding2021differentially}). Here is an example: Given $\Delta>0$, the quantizer $Q(x)$ with the following probability distribution satisfies Assumption \ref{assumption 3: quantizer} by Lemma 1 of \cite{aysal2008distributed}:\vspace{-0.5em}
		\begin{align}\label{quantizer mainly used}
			\begin{cases}
				\p\left(\left.Q(x)=\Delta\lfloor \frac{x}{\Delta} \rfloor\right|x\right)=1-\frac{x}{\Delta}+\lfloor \frac{x}{\Delta} \rfloor;\cr
				\p\left(\left.Q(x)=\Delta\left(\lfloor \frac{x}{\Delta} \rfloor+1\right)\right|x\right)=\frac{x}{\Delta}-\lfloor \frac{x}{\Delta} \rfloor.
			\end{cases}
	\end{align}
\end{rmk}\vspace{-1.5em}
\subsection{Differential privacy}\label{sec2d}\vspace{-0.3em}
As shown in \cite{wang2023quantization,wang2023decentralized}, there are two kinds of adversary models widely used in the privacy-preserving issue for distributed stochastic optimization:
\begin{itemize}[leftmargin=*,itemsep=0em]
	\item A \emph{semi-honest} adversary. This kind of adversary is defined as a node within the network which has access to certain internal states (such as $x_{i,k}$ from Node $i$), follows the prescribed protocols and accurately computes iterative state correctly. However, it aims to infer the sensitive information of other nodes.
	\item An \emph{eavesdropper}. This kind of adversary refers to an external adversary who has capability to  wiretap and monitor all communication channels, allowing them to capture distributed messages from any node. This enables the eavesdropper to infer the sensitive information of internal nodes.
\end{itemize}
When solving the empirical risk minimization problem \eqref{question of interest 2}, the stochastic first-order oracle needs data samples to return sampled gradients. Meanwhile, the adversaries above infer the sensitive information of data samples from sampled gradients (\!\!~\cite{zhu2019deep}). Inspired by \cite{wang2022differentially1,liu2024distributed}, a symmetric binary relation called \emph{adjacency relation} is defined as follows:
\begin{def0}\label{def1:adjacency relation}
	(Adjacency relation) Let $\mathcal{D}$$=$$\{\xi_{i,l}:i\in\mathcal{V},l=1,\dots,D\}$, $\mathcal{D}^\prime\hspace{-0.2em}=\hspace{-0.2em}\{\xi_{i,l}^{\prime}:i\hspace{-0.2em}\in\hspace{-0.2em}\mathcal{V},l=1,\dots,D\}$ be two sets of data samples. If there exists $C>0$ and exactly one pair of data samples $\xi_{i_0,l_0},\xi_{i_0,l_0}^\prime$ in $\mathcal{D},\mathcal{D}^\prime$ such that for any $x\in\mathbb{R}^{r}$,\vspace{-0.7em}
		\begin{align}\label{eq: in definition 3}
			\begin{cases}
				\!\|\nabla\ell_i(x,\xi_{i,l})\!-\!\nabla\ell_i(x,\xi_{i,l}^\prime)\|\!\leq\!C,&\hspace{-0.8em}\text{ if }i=i_0\text{ and }l=l_0;\cr
				\noalign{\vskip -0.3em}
				\!\|\nabla\ell_i(x,\xi_{i,l})\!-\!\nabla\ell_i (x,\xi_{i,l}^\prime)\|\!=\!0,&\hspace{-0.8em}\text{ if }i\neq i_0\text{ or }l\ne l_0,
			\end{cases}\!\!\!\!
		\end{align}{\vskip -6pt}\noindent then $\mathcal{D}$ and $\mathcal{D}^\prime$ are said to be adjacent, denoted by $\text{Adj}(\mathcal{D},\mathcal{D}^\prime)$.
\end{def0}
\begin{rmk}
	The constant $C$ characterizes the ``closeness'' of a pair of data samples $\xi_{i_0,l_0}$, $\xi_{i_0,l_0}^\prime$. By \eqref{eq: in definition 3}, the larger the constant $C$ is, the larger the allowed magnitude of sampled gradients between adjacent datasets is, and thus, the better the privacy protection level is. Moreover, for any given constant $C$, as long as there exists a pair of sample sets $\mathcal{D},\mathcal{D}^\prime$ satisfying the adjacency relation defined by this constant $C$, then the privacy analysis in Section \ref{subsection 3.2} holds for $\text{Adj}(\mathcal{D},\mathcal{D}^\prime)$.
\end{rmk}
\begin{rmk}
	Definition \ref{def1:adjacency relation} allows us to avoid the assumption of bounded gradients required in \cite{zhang2018improving,li2018differentially,huang2019dp,gratton2021privacy,liu2024distributed,xu2022dp,wang2023decentralized,ding2021differentially,wang2023quantization} to achieve differential privacy. Specifically, since $\mathcal{D},\mathcal{D}^\prime$ have finite data samples, it follows that $\max_{\omega\in\mathcal{D}\cup\mathcal{D}^\prime}\|\omega\|$$<$$\infty$. Then, for any $C\geq2 L_2$ $\max_{\omega\in\mathcal{D}\cup\mathcal{D}^\prime}\|\omega\|$ and $x\in\mathbb{R}^r$, by Assumption \ref{assumption 2: cost functions}(i), we have\vspace{-0.6em}
	\begin{align*}
	\begin{cases}
		{\footnotesize \begin{matrix}
		\|\nabla\ell_i(x,\xi_{i,l})-\nabla\ell_i(x,\xi_{i,l}^\prime)\|\leq  L_2\|\xi_{i,l}-\xi_{i,l}^\prime\|\\
		\leq2 L_2\max_{\omega\in\mathcal{D}\cup\mathcal{D}^\prime}\|\omega\|\leq C,~~~~~~~~~~~~
		\end{matrix}}&\hspace{-0.8em}\text{ if }i=i_0\text{ and }l=l_0;\cr
		\|\nabla\ell_i(x,\xi_{i,l})-\nabla\ell_i (x,\xi_{i,l}^\prime)\|=0,&\hspace{-0.8em}\text{ if }i\neq i_0\text{ or }l\ne l_0.
	\end{cases}
	\end{align*} {\vskip -4pt}\noindent This shows \eqref{eq: in definition 3} holds for any $x\in\mathbb{R}^r$. Thus, \eqref{eq: in definition 3} holds no matter whether gradients are bounded or not.
\end{rmk}

To give the privacy-preserving level of the algorithm, we adopt the definition of $(\epsilon,\delta)$-differential privacy as follows:
\begin{def0}\label{def2:differential privacy}
	(\!\cite{liu2024distributed}) ($(\epsilon,\delta)$-differential privacy) Given $\epsilon>0$,$0$$<$$\delta$$\leq$$1$, a mechanism $\mathcal{M}$ achieves $(\epsilon,\delta)$-differential privacy for $\text{Adj}(\mathcal{D},\mathcal{D}^\prime)$ if  $\p(\mathcal{M}(\mathcal{D})\!\in\!\mathcal{T})\!\leq\!e^{\epsilon}\p(\mathcal{M}(\mathcal{D}^\prime)\!\in\!\mathcal{T})+\delta$ for any Borel-measurable set $\mathcal{T}$$\subset$$\text{Range}(\mathcal{M})$.
\end{def0}
\vspace{-1em}
\section{Main result}\label{section3}\vspace{-0.5em}
\subsection{The proposed algorithm}\vspace{-0.5em}
In this subsection, we give a differentially private distributed nonconvex stochastic optimization algorithm with quantized communication. The detailed implementation steps are given in Algorithm \ref{algorithm1}.\vspace{-0.6em}
\restylefloat{algorithm}
\begin{algorithm}[H]
	\caption{\fontsize{9.7}{10}\selectfont Differentially private distributed nonconvex stochastic optimization algorithm with quantized communication}
	\label{algorithm1}
	%%%%%%%%%%%%%%%%%%%%%%%%%%%%%%%%%%%%%%%%%%%%%%%%%%%%%%%%%%%%%%
	\renewcommand{\algorithmicrequire}{\textbf{Parameters:}}
	\renewcommand{\algorithmicensure}{\textbf{Initialization:}}
	\begin{algorithmic}[1]
		\Ensure $x_{i,0}$$\in$$\mathbb{R}^{r}$, weight matrix $(a_{ij})_{1\leq i,j\leq n}$, iteration maximum $T$, step-sizes $\alpha_T$$=$$\frac{a_1}{(T+1)^{u}}$, $\beta_T$$=$$\frac{a_2}{(T+1)^{v}}$ and sample-size $\gamma_T$$=$$\lfloor a_3 T^{s}\rfloor+1$.
		
		\hspace{-3.7em}{\bf for} $k=0,\dots, T$, {\bf do}
		\State Node $i$ adds noise $d_{i,k}$ to $x_{i,k}$ and computes the quantized information $z_{i,k}$$=$$Q$$($$x_{i,k}$$+$$d_{i,k}$$)$$=$$[$$Q$$($$x_{i,k}^{(1)}$$+$$d_{i,k}^{(1)}$$)$,$\dots$,  $Q$$($$x_{i,k}^{(r)}$$+$$d_{i,k}^{(r)}$$)$$]^\top$ with the probabilistic quantizer in the form of \eqref{quantizer mainly used}, where $d_{i,k} \sim N(0,\sigma_k^2 I_{r})$.
		\State Node $i$ broadcasts $z_{i,k}$ to its neighbors $j\in\mathcal{N}_{i}$, receives $z_{j,k}$ from its neighbors $j$$\in $$\mathcal{N}_{i}$, and aggregates the received information by\vspace{-0.55em}
		\begin{align}\label{eq5}
			\tilde{x}_{i,k}=(1-\beta_T)x_{i,k}+\beta_T\sum_{j\in \mathcal{N}_{i}}a_{ij}z_{j,k}.
		\end{align}\vspace{-1.2em}
		\State Node $i$ takes $\gamma_T$ different data samples $\zeta_{i,k,1}$,$\dots$,$\zeta_{i,k,\gamma_T}$ uniformly from $\mathcal{D}_i$ simultaneously (i.e., without replacement) to generate sampled gradients $\nabla\ell_i(x_{i,k},\zeta_{i,k,1})$,$\dots$, $\nabla\ell_i(x_{i,k},\zeta_{i,k,\gamma_T})$. Then, Node $i$ puts these data samples back into $\mathcal{D}_i$.
		\State Node $i$ computes the averaged sampled gradient by\vspace{-0.55em}
		\begin{align}\label{eq:generate gradient}
			\nabla\ell_{i,k}=\frac{1}{\gamma_T}\sum_{l=1}^{\gamma_T} \nabla\ell_i(x_{i,k},\zeta_{i,k,l}).
		\end{align}\vspace{-1em}
		\State Node $i$ updates its state by\vspace{-0.8em}
		\begin{align}\label{algorithm: per-node}
			x_{i,k+1}=\tilde{x}_{i,k}-\alpha_T\nabla\ell_{i,k}.
		\end{align}{\vskip -8pt}
		\hspace{-3.7em}{\bf end for}
	\end{algorithmic}
\end{algorithm}\aftergroup\ignorespacesafterend
\vspace{-1em}
%\parfillskip=-5pt\par
%\setlength{\belowdisplayskip}{-5pt}
%\makeatletter
%\def\algocf@afterblock{\vskip-3em} 
%\makeatother
\begin{rmk}
	By the subsampling method in Step 3 of Algorithm \ref{algorithm1}, there are sufficient data samples to run Algorithm 1 since data samples are put back into the dataset $\mathcal{D}_i$ at each iteration. Specially, when each node only has one data sample (i.e., $D=1$), let $s=0$, $a_3=\frac{1}{2}$. Then, the sample-size $\gamma_T=1=D$. In this case, Algorithm \ref{algorithm1} still works.
\end{rmk}\vspace{-1.3em}
\subsection{Privacy analysis}\label{subsection 3.2}\vspace{-0.3em}
In this subsection, we will show the differential privacy analysis of  Algorithm \ref{algorithm1}. Inspired by \cite{wang2022differentially1}, we first provide the sensitivity of the algorithm, which helps us to analyze the differential privacy of the algorithm.
\begin{def0}\label{def3}
	(Sensitivity) Given $\text{Adj}(\mathcal{D},\mathcal{D}^\prime)$, and a query $q$. For any $k=0,\dots, T$, let $\mathcal{D}_{k}$$=$$\{\zeta_{i,k,l}$$:$$ i\in\mathcal{V},l=1,\dots,\gamma_T\}$, $\mathcal{D}_{k}^\prime$$=$$\{\zeta_{i,k,l}^{\prime}$$:$$i\!\in\!\mathcal{V},l=1,\dots,\gamma_T\}$ be the data samples taken from $\mathcal{D},\mathcal{D}^\prime$ at the $k$-th iteration, respectively. Then,  the sensitivity of  Algorithm \ref{algorithm1} at the $k$-th iteration  is  defined as follows:\vspace{-0.3em}
	\begin{align}\label{eq:definition of sensitivity,1}
			\Delta_{k}^{q}\triangleq\!\!\sup_{\mathcal{S}\in\mathbb{R}^{nr}}\sup_{y\in\mathcal{S}}\sup_{\text{Adj}(\mathcal{D},\mathcal{D}^\prime)}\!\!\|q(\mathcal{D}_{k}|z_k \!=\! y)\!-\!q(\mathcal{D}_{k}^\prime|z_k^\prime\!=\! y)\|.
	\end{align}
\end{def0}
\begin{rmk}\label{rmk4}
	Definition \ref{def3} captures the magnitude by which one node's data sample can change the query $q$ in the worst case. The sensitivity $\Delta_{k}^{q}$ is commonly used in \cite{zhang2018improving,li2018differentially,huang2019dp,ding2021differentially,gratton2021privacy,xu2022dp,wang2023decentralized,liu2024distributed}, and determines how much noise should be added at the $k$-th iteration to achieve $(\epsilon_k,\delta_k)$-differential privacy. In Algorithm \ref{algorithm1}, the query $q(\mathcal{D}_{k}|z_k\!\!=\!\!y)$ denotes the state $x_{k+1}$ at the $k$-th iteration under data samples $\mathcal{D}_{k}$ and the execution $z_k$$=$$[z_{1,k}^\top,\dots,z_{n,k}^\top]^\top$$=$$[y_1^\top,\dots,y_n^\top]^\top$$=$$y$, i.e., $q(\mathcal{D}_{k}|z_k\!=\!y)$ $=$$x_{k\!+\!1}$$=$$[x_{1,k\!+\!1}^\top,\dots,x_{n,k\!+\!1}^\top]^\top$. The mechanism $\mathcal{M}(\mathcal{D}_{k})$ denotes the quantized noise-perturbed state at the $k$-th iteration, i.e., $\mathcal{M}(\mathcal{D}_{k})$$=$$Q(q(\mathcal{D}_k|z_k\!=\!y)$$+$$d_{k\!+\!1})$$=$$Q(x_{k\!+\!1}\!+\!d_{k\!+\!1})$$=$ $[Q(x_{1,k\!+\!1}$$+$$d_{1,k\!+\!1})^{\!\top},\dots,Q(x_{n,k\!+\!1}$$+$$d_{n,k\!+\!1})^{\!\top}]^{\!\top}$$=$$z_{k\!+\!1}$.
\end{rmk}
The following lemma gives the sensitivity $\Delta_{k}$ of Algorithm~\ref{algorithm1} for any $k=0,\dots, T$.
\begin{lemma}\label{lemma1: sensitivity}
	At the $k$-th iteration, the sensitivity of Algorithm~\ref{algorithm1} satisfies $\Delta_{k}^{q}\leq \frac{\alpha_T C}{\gamma_T}\left(\sum_{m=0}^{k}|1-\beta_T|^m\right)$.
\end{lemma}\vspace{0.3em}
{\bf Proof:} When $k=0$, \eqref{eq:definition of sensitivity,1} can be written as\vspace{-0.6em}
\begin{align}\label{eq:sensitivity k=0 ,1}
		\Delta_0^{q}\!=&\!\!\sup_{\mathcal{S}\in\mathbb{R}^{nr}}\sup_{y\in\mathcal{S}}\sup_{\text{Adj}(\mathcal{D},\mathcal{D}^\prime)}\!\left\|q(\mathcal{D}_{0}|z_0 \!=\! y)\!-\!q(\mathcal{D}_{0}^\prime|z_0^\prime\!=\! y)\right\|\cr
		=&\!\!\sup_{\mathcal{S}\in\mathbb{R}^{nr}}\sup_{y\in\mathcal{S}}\sup_{\text{Adj}(\mathcal{D},\mathcal{D}^\prime)}\!\left\|x_1\!-\!x_1^{\prime}\right\|.
\end{align}{\vskip -6pt}

From \eqref{eq:sensitivity k=0 ,1}, it can be seen that $z_{i,0}$$=$$y_i$$=$$z_{i,0}^\prime$ holds for any $i$$\in$$\mathcal{V}$. Moreover, since $x_{i,0}=x_{i,0}^\prime$ holds for any  $i\in\mathcal{V}$, by \eqref{eq5}, $\tilde{x}_{i,0}=\tilde{x}_{i,0}^{\prime}$ holds for any $i\in\mathcal{V}$. Let $\nabla\ell_0=[\nabla\ell_{1,0}^\top,\dots,\nabla\ell_{n,0}^\top]^\top$. Then, substituting \eqref{algorithm: per-node} into \mbox{\eqref{eq:sensitivity k=0 ,1} implies}\vspace{-0.5em}
\begin{align}\label{eq:sensitivity k=0 ,3}
\Delta_0^{q}=&\sup_{\mathcal{S}\in\mathbb{R}^{nr}}\sup_{y\in\mathcal{S}}\sup_{\text{Adj}(\mathcal{D},\mathcal{D}^\prime)}\left\|\alpha_T(\nabla\ell_0-\nabla\ell_0^{\prime})\right\|\cr
\noalign{\vskip -3pt}
	=&\sup_{\text{Adj}(\mathcal{D},\mathcal{D}^\prime)}\left\|\alpha_T(\nabla\ell_0-\nabla\ell_0^{\prime})\right\|.
\end{align}
{\vskip -6pt}\noindent By Definition \ref{def1:adjacency relation}, since $\mathcal{D}$ and $\mathcal{D}^\prime$ are adjacent, there exists exactly one pair of data samples $\xi_{i_0,l_0},\xi_{i_0,l_0}^{\prime}$ in $\mathcal{D}$ and $\mathcal{D}^\prime$ such that \eqref{eq: in definition 3} holds. This implies that $\nabla\ell_{j,0}=\nabla\ell_{j,0}^\prime$ holds for any node $j\neq i_0$. Thus, \eqref{eq:sensitivity k=0 ,3} can be rewritten as\vspace{-0.5em}
\begin{align}\label{eq:sensitivity k=0 ,4}
	\smash{\Delta_0^{q}=\alpha_T\sup_{\text{Adj}(\mathcal{D},\mathcal{D}^\prime)}\|\nabla\ell_{i_0,0}-\nabla\ell_{i_0,0}^\prime\|.}
\end{align}\indent Since $\gamma_T$ different data samples are taken uniformly from $\mathcal{D}$, $\mathcal{D}^\prime$ simultaneously, there exists at most one pair of data samples $\zeta_{i_0,0,l_1},\zeta_{i_0,0,l_1}^\prime$ such that $\zeta_{i_0,0,l_1}=\xi_{i_0,l_0}$, $\zeta_{i_0,0,l_1}^\prime=\xi_{i_0,l_0}^\prime$. Thus, by \eqref{eq:generate gradient}, \eqref{eq:sensitivity k=0 ,4} can be rewritten as{\vskip -15pt}
\begin{align*}
	\Delta_0^{q}=&\frac{\alpha_T}{\gamma_T}\!\!\sup_{\text{Adj}(\mathcal{D},\mathcal{D}^\prime)}\!\|\!\sum_{l=1}^{\gamma_T}\!(\nabla\ell_{i_0}(x_{i_0,0},\zeta_{i_0,0,l})\!-\! \nabla\ell_{i_0}(x_{i_0,0},\zeta_{i_0,0,l}^{\prime}))\!\|\\
	\noalign{\vskip -2pt}
	=&\frac{\alpha_T}{\gamma_T}\!\!\sup_{\text{Adj}(\mathcal{D},\mathcal{D}^\prime)}\!\left\|\nabla\ell_{i_0}(x_{i_0,0},\zeta_{i_0,0,l_1})\!-\!\nabla\ell_{i_0}(x_{i_0,0},\zeta_{i_0,0,l_1}^{\prime})\right\|\\
	\noalign{\vskip -4pt}
	\leq&\frac{\alpha_T}{\gamma_T}\!\!\sup_{\text{Adj}(\mathcal{D},\mathcal{D}^\prime)}\!\left\|\nabla\ell_{i_0}(x_{i_0,0},\xi_{i_0,l_0})\!-\!\nabla\ell_{i_0}(x_{i_0,0},\xi_{i_0,l_0}^{\prime})\right\|\!\!\cr
	\leq& \frac{\alpha_TC}{\gamma_T}.
\end{align*}{\vskip -4pt}\indent When $k=0,\dots, T$, by \eqref{eq:definition of sensitivity,1} we have\vspace{-0.6em}
\begin{align}\label{eq:sensitivity iterative, k>0,1}
	\Delta_{k}^{q}=&\!\!\sup_{\mathcal{S}\in\mathbb{R}^{nr}}\sup_{y\in\mathcal{S}}\sup_{\text{Adj}(\mathcal{D},\mathcal{D}^\prime)}\!\left\|q(\mathcal{D}_{k}|z_k \!=\! y)\!-\!q(\mathcal{D}_{k}^\prime|z_k^\prime\!=\! y)\right\|\cr
	=&\!\!\sup_{\mathcal{S}\in\mathbb{R}^{nr}}\sup_{y\in\mathcal{S}}\sup_{\text{Adj}(\mathcal{D},\mathcal{D}^\prime)}\left\|x_{k+1}-x_{k+1}^\prime\right\|.
\end{align}
{\vskip -5pt}\noindent From \eqref{eq:sensitivity iterative, k>0,1}, it can be seen that $z_{i,k}$$=$$z_{i,k}^\prime$ holds for any $i$$\in$$\mathcal{V}$, $k$$=$$0$$,$$\dots$$, $$T$, and thus, $z_{i,T}$$=$$z_{i,T}^\prime$ holds for any $i$$\in$$\mathcal{V}$. Moreover, note that $x_{i,0}$$=$$x_{i,0}^\prime$ holds for any $i$$\in$$\mathcal{V}$ and $\nabla\ell_{j,m}$$=$$\nabla\ell_{j,m}^{\prime}$ holds for any node $j$$\neq$$i_0$, $m$$=$$0$$,$$\dots$$, $$k$. Then, by \eqref{eq5}, $\tilde{x}_{j,k}$$=$$\tilde{x}_{j,k}^{\prime}$ holds for any node $j$$\neq$$i_0$. Thus, by \eqref{algorithm: per-node}, $x_{j,k+1}$$=$$x_{j,k+1}^{\prime}$ holds for any node $j$$\neq$$i_0$. Hence, \eqref{eq:sensitivity iterative, k>0,1} can be rewritten as\vspace{-0.78em}
\begin{align}\label{eq:sensitivity iterative, k>0,1.5}
	\Delta_{k}^{q}=\!\!\sup_{\mathcal{S}\in\mathbb{R}^{nr}}\sup_{y\in\mathcal{S}}\sup_{\text{Adj}(\mathcal{D},\mathcal{D}^\prime)}\|x_{i_0,k+1}-x_{i_0,k+1}^\prime\|.
\end{align}{\vskip -5pt}
\noindent Note that $z_{i_0,k}$$=$$z_{i_0,k}^\prime$. Then, substituting \eqref{eq5}-\eqref{algorithm: per-node} into \eqref{eq:sensitivity iterative, k>0,1.5} implies\vspace{-0.6em}
\begin{align}\label{eq:sensitivity iterative, k>0,2}
	\Delta_{k}^{q}=&\!\!\sup_{\mathcal{S}\in\mathbb{R}^{nr}}\sup_{y\in\mathcal{S}}\sup_{\text{Adj}(\mathcal{D},\mathcal{D}^\prime)}\| (\tilde{x}_{i_0,k}\!-\!\tilde{x}_{i_0,k}^{\prime})\!-\!\alpha_T(\nabla\ell_{i_0,k}\!-\!\nabla\ell_{i_0,k}^{\prime})\|\nonumber\\
	=&\!\!\sup_{\text{Adj}(\mathcal{D},\mathcal{D}^\prime)}\| (\tilde{x}_{i_0,k}\!-\!\tilde{x}_{i_0,k}^{\prime})\!-\!\alpha_T(\nabla\ell_{i_0,k}\!-\!\nabla\ell_{i_0,k}^{\prime})\|\cr
	\noalign{\vskip -2pt}
	\leq&\!\!\sup_{\text{Adj}(\mathcal{D},\mathcal{D}^\prime)}\!\!\|(1-\beta_T)(x_{i_0,k}-x_{i_0,k}^{\prime})\|\cr
	\noalign{\vskip -7pt}
	&\!\!+\!\!\!\!\!\!\sup_{\text{Adj}(\mathcal{D},\mathcal{D}^\prime)}\!\!\!\!\!\|\frac{\alpha_T}{\gamma_T}\!\sum_{l=1}^{\gamma_T}\!(\!\nabla\ell_{i_{0}}\!(x_{i_{0}\!,k},\zeta_{i_{0}\!,k\!,l})\!\!-\!\!\nabla\ell_{i_{0}}\!(x_{i_{0}\!,k},\zeta_{i_{0}\!,k\!,l}^{\prime})\!)\!\|\!.
\end{align}

{\vskip -6pt}\indent Since $\mathcal{D}$ and $\mathcal{D}^\prime$ are adjacent, there exists at most one pair of data samples $\zeta_{i_0,k,l_{k\!+\!1}}$, $\zeta_{i_0,k,l_{k\!+\!1}}^{\prime}$ such that $\zeta_{i_0,k,l_{k\!+\!1}}=\xi_{i_0,l_0}$, $\zeta_{i_0,k,l_{k\!+\!1}}^{\prime}=\xi_{i_0,l_0}^\prime$. Then, \eqref{eq:sensitivity iterative, k>0,2} can be rewritten as\vspace{-0.3em}
\begin{align}\label{eq:sensitivity iterative}
		\Delta_{k}^{q}\leq&\sup_{\text{Adj}(\mathcal{D},\mathcal{D}^\prime)}\left\|(1-\beta_T)(x_{i_0,k}-x_{i_0,k}^{\prime})\right\|\cr
		\noalign{\vskip -2pt}
		&+\frac{\alpha_T}{\gamma_T}\!\!\sup_{\text{Adj}(\mathcal{D},\mathcal{D}^\prime)}\!\!\left\|\nabla\ell_{i_0}(x_{i_0,k},\xi_{i_0,l_0})\!-\!\nabla\ell_{i_0}(x_{i_0,k},\xi_{i_0,l_0}^{\prime})\right\|\cr
		\noalign{\vskip -5pt}
		\leq&|1-\beta_T|\sup_{\text{Adj}(\mathcal{D},\mathcal{D}^\prime)}\left\|x_{i_0,k}-x_{i_0,k}^{\prime}\right\|+\frac{\alpha_TC}{\gamma_T}.
\end{align}
{\vskip -5pt}\indent By iteratively computing \eqref{eq:sensitivity iterative}, this lemma is proved. \hfill$\blacksquare$

Next, we show that Algorithm \ref{algorithm1} achieves $(\epsilon_k,\delta_k)$-differential privacy at the $k$-th iteration for any $k=0,\dots, T$.
%from Theorem A.1 in \cite{dwork2014algorithmic}, which helps us to achieve  by the Gaussian mechanism.
\begin{lemma}\label{lemma2: gaussian mechanism}
	Given $\text{Adj}(\mathcal{D}, \mathcal{D}^\prime)$, the query $q$ and $\epsilon_k$$>$$0$, $0$$<$$\delta_k$$\leq$$1$, $x_{k+1},x_{k+1}^\prime$$\in$$\mathbb{R}^{nr}$ for any $k=0,\dots, T$. Then, for any Borel-measurable set $\mathcal{S}$$\subset$$\mathbb{R}^{nr}$, the mechanism $\mathcal{M}$$($$\mathcal{D}_{k}$$)$$=$ $Q$$($$x_{k\!+\!1}$$+$$d_{k\!+\!1}$$)$ satisfies\vspace{-0.5em}
	\begin{align*}
		&\p(\mathcal{M}(\mathcal{D}_{k})\in\mathcal{S})\leq e^{\epsilon_k}\p(\mathcal{M}(\mathcal{D}_{k}^\prime)\in\mathcal{S})+\delta_k,
	\end{align*}{\vskip -4pt}\noindent where $d_{k\!+\!1}\!\!=\!\![d_{1,k\!+\!1}^\top,\dots,d_{n,k\!+\!1}^\top]^\top\!\!\!\sim\!\!N(0,\sigma_{k+1}^2\!I_{nr})$ is a Gaussian noise with the variance $\sigma_{k+1}^2=4\ln\left(\frac{1.25}{\delta_k}\right)\left(\frac{\Delta_{k}^{q}}{\epsilon_k}\right)^2$.
\end{lemma}\vspace{0.1em}
{\bf Proof.} Note that Gaussian noises $d_{k+1}$, $d_{k+1}^\prime$ have the variance $\sigma_{k+1}^2$$=$$4\ln(\frac{1.25}{\delta_k})(\frac{\Delta_k^{q}}{\epsilon_k})^2$ for any $k=0,\dots, T$. Then, for any Borel-measurable $\mathcal{O}$$\subset$$\mathbb{R}^{nr}$, by the Gaussian mechanism \cite[Th. A.1]{dwork2014algorithmic} \mbox{we have}\vspace{-0.4em}
\begin{align}\label{eq6}
	\p(x_{k+1}+d_{k+1}\!\in\!\mathcal{O})\leq e^{\epsilon_k}\p(x_{k+1}^\prime+d_{k+1}^\prime\!\in\!\mathcal{O})+\delta_k.
\end{align}
{\vskip -4pt}\noindent Thus, for the Borel-measurable set $\mathcal{S}$$=$$\mathcal{M}(\mathcal{O})$, by \eqref{eq6} and the post-processing property \cite[Prop. 2.1]{dwork2014algorithmic} we have $\p(\mathcal{M}(\mathcal{D}_{k})\in\mathcal{S})\leq e^{\epsilon_k}\p(\mathcal{M}(\mathcal{D}_{k}^\prime)\in\mathcal{S})+\delta_k. $  \hfill$\blacksquare$
\begin{lemma}\label{lemma3}
		\hspace{-0.05em}\cite[Cor. B.2]{dwork2014algorithmic} Given $\text{Adj}$$($$\mathcal{D}$$,$$\mathcal{D}^\prime$$)$, if the mechanism $\mathcal{M}$$($$\mathcal{D}_k$$)$ achieves $($$\epsilon_k$$,$$\delta_k$$)$-differential privacy for any $k$$=$$0$$,$$\dots$$,$$ T$, then the mechanism $\mathcal{M}$$(\mathcal{D})$$=$$($$\mathcal{M}$$($$\mathcal{D}_0$$)$, $\dots$, $\mathcal{M}$$($$\mathcal{D}_T$$)$$)$ achieves  $($$\sum_{k=0}^{T}\epsilon_k$,$\sum_{k=0}^{T}\delta_k$$)$-differential privacy.
\end{lemma}\vspace{0.3em}
\begin{thm0}\label{thm1}
For any $T=0,1,\dots,k=0,\dots, T$, let\vspace{-0.5em}
\begin{align*}
	&\alpha_T=\frac{a_1}{(T+1)^{u}},\ \beta_T=\frac{a_2}{(T+1)^{v}},\ 
	\gamma_T=\lfloor a_3T^{s}\rfloor+1,\cr &\sigma_k=(k+1)^{w},\delta_k=\frac{1}{(k+2)^{t}},\ a_1, \ a_2, \ a_3>0.
\end{align*}  
{\vskip -5pt}\noindent If $0<a_2<1$ and $t>0$, then Algorithm \ref{algorithm1} achieves $(\epsilon,\delta)$-differential privacy over finite iterations $T$, where{\vskip -5pt}
\begin{align}\label{eq: epsilon, delta}
	\epsilon=&\sum_{k=0}^{T}\epsilon_k\leq\sum_{k=0}^{T}\frac{2C a_1\sqrt{\ln (1.25(k+2)^t)}}{a_2(T+1)^{u-v}(\lfloor a_3T^s\rfloor+1)(k+2)^w},\notag\\
	\delta=&\sum_{k=0}^{T}\delta_k=\sum_{k=0}^{T}\frac{1}{(k+2)^t}.
\end{align}
{\vskip -5pt}\noindent Furthermore, if $u+s-v$$>$$\max\{1-w,0\}$, $t\geq 2$, then Algorithm \ref{algorithm1} achieves finite cumulative differential privacy budgets $\epsilon$, $\delta$ over infinite iterations.
\end{thm0}

{\bf Proof.} By Lemma \ref{lemma2: gaussian mechanism}, the mechanism $\mathcal{M}(\mathcal{D}_k)$ achieves $(\epsilon_k,\delta_k)$-differential privacy with $\epsilon_k$$=$$2\sqrt{\ln (\frac{1.25}{\delta_k})}\frac{\Delta_k^{q}}{\sigma_{k+1}}$ for any $k$$=$$0,$$\dots$$, $$T$. Then, using Lemma \ref{lemma3}, it can be seen that the mechanism $\mathcal{M}(\mathcal{D})$ achieves the $(\sum_{k=0}^{T}\epsilon_k$,$\sum_{k=0}^{T}\delta_k)$-differential privacy, i.e., $\p(\mathcal{M}(\mathcal{D})\!\in\!\mathcal{T})\!\leq\!e^{\sum_{k=0}^{T}\epsilon_k}\p(\mathcal{M}(\mathcal{D}^\prime)\!\in\!\mathcal{T})+\sum_{k=0}^{T}\delta_k$ for any Borel-measurable set $\mathcal{T}\subset\text{Range}(\mathcal{M})=\mathbb{R}^{n(T+1)r}$.
{\vskip 2pt}\indent By Lemma \ref{lemma1: sensitivity}, the cumulative differential privacy budget $\sum_{k=0}^{T}\epsilon_k$ can be rewritten as\vspace{-0.6em}
\begin{align}\label{eq7}
	\sum_{k=0}^{T}\epsilon_k=&\sum_{k=0}^{T}\frac{2\alpha_T C\sqrt{\ln (\frac{1.25}{\delta_k})}\Delta_k^{q}}{\sigma_{k+1}}\cr
	\noalign{\vskip -7pt}
	\leq&\sum_{k=0}^{T}\frac{2\alpha_T C\sqrt{\ln (\frac{1.25}{\delta_k})}(\sum\limits_{m=0}^{k}|1-\beta_T|^m)}{\gamma_T\sigma_{k+1}}.
\end{align}{\vskip -4pt}\noindent Since $0$$<$$a_2$$<$$1$, it can be seen that $0$$<$$\beta_T$$<$$1$. Then, we have $\sum_{m=0}^{k}$$|1$$-$$\beta_T|^m$$=$$\frac{1-(1-\beta_T)^{k+1}}{\beta_T}$$\leq$$\frac{1}{\beta_T}$. Substituting it into \mbox{\eqref{eq7} implies}\vspace{-0.7em}
\begin{align*}
	\sum_{k=0}^{T}\epsilon_k\leq&\sum_{k=0}^{T}\frac{2\alpha_T C\sqrt{\ln (\frac{1.25}{\delta_k})}}{\beta_T\gamma_T\sigma_{k+1}}\cr
	=&\sum_{k=0}^{T}\frac{2C a_1\sqrt{\ln (1.25(k+2)^t)}}{a_2(T+1)^{u-v}(\lfloor a_3T^s\rfloor+1)(k+2)^w}\cr
	\leq&\sum_{k=0}^{T}\frac{2Ca_1\sqrt{\ln (1.25 (T+2)^{t})}}{a_2(a_3 T^{u+s-v}+1)(k+2)^{w}}\cr
	=&\frac{2Ca_1\sqrt{\ln (1.25 (T+2)^{t})}}{a_2(a_3 T^{u+s-v}+1)}\sum_{k=0}^{T}\frac{1}{(k+2)^{w}}\cr
	\noalign{\vskip -4pt}
	=&O(\frac{(\ln(T\!+\!2))^{\frac{3}{2}}}{(T+1)^{u+s-v-\max\{1-w,0\}}}).
\end{align*}
{\vskip -6pt}\noindent Thus, if $u+s-v\!>\!\max\{1-w,0\}$, then the cumulative differential privacy budget $\sum_{k=0}^{T}\epsilon_k$ is finite even over infinite iterations. In addition, if $t\geq 2$, then the cumulative differential privacy budget $\sum_{k=0}^{T}\delta_k$ is finite even over infinite iterations. Hence, this theorem is proved. \hfill$\blacksquare$
\begin{rmk}
	By Theorem \ref{thm1}, $(\epsilon,\delta)$-differential privacy is achieved for all nodes. When $\mathcal{T}=\mathbb{R}^{(i_0-1)(T+1)r}\times\mathcal{S}\times\mathbb{R}^{(n-i_0)(T+1)r}$ for any Borel-measurable set $\mathcal{S}\in\mathbb{R}^{(T+1)r}$, we have\vspace{-0.5em}
\begin{align*}		
	&\p((z_{i_0,0},\dots,z_{i_0,T})\in\mathcal{S})\cr
	=&\p((z_{1,0},\dots,z_{1,T},\dots,z_{i_0,0},\dots,z_{i_0,T},\dots,z_{n,0},\dots,z_{n,T})\cr
	&\in\mathbb{R}^{(i_0-1)(T+1)r}\times\mathcal{S}\times\mathbb{R}^{(n-i_0)(T+1)r})\cr
	\leq&e^{\epsilon}\p((z_{1,0}^\prime,\dots,z_{1,T}^\prime,\dots,z_{i_0,0}^\prime,\dots,z_{i_0,T}^\prime,\dots,z_{n,0}^\prime,\dots,z_{n,T}^\prime)\cr
	&\in\mathbb{R}^{(i_0-1)(T+1)r}\times\mathcal{S}\times\mathbb{R}^{(n-i_0)(T+1)r})+\delta\cr
	=&e^{\epsilon}\p((z_{i_0,0}^\prime,\dots,z_{i_0,T}^\prime)\in\mathcal{S})+\delta.
\end{align*}
{\vskip -5pt}\noindent This implies $(\epsilon,\delta)$-differential privacy is achieved for a particular node~$i_0$. In this case, the sensitive information of all nodes can be protected against both the eavesdropper and the semi-honest adversary. Thus, Theorem~\ref{thm1} provides a unified privacy analysis framework for both adversary models presented in Subsection~\ref{sec2d}.
\end{rmk}
\begin{rmk}
	Theorem \ref{thm1} shows how step-size parameters $u$, $v$, the sample-size parameter $s$ and the privacy noise parameter $w$ affect cumulative differential privacy budgets $\epsilon$, $\delta$. As shown in \eqref{eq: epsilon, delta}, the larger the step-size parameter $u$, the sample-size parameter $s$ and the privacy noise parameter $w$ are, the smaller cumulative differential privacy budgets $\epsilon$, $\delta$ are. In addition, the smaller the step-size parameter $v$ is, the smaller cumulative differential privacy budgets $\epsilon$, $\delta$ are.
\end{rmk}
\begin{rmk}
	By \eqref{eq: epsilon, delta}, the larger the sample-size $\gamma_T$ is, the smaller cumulative differential privacy budgets $\epsilon$, $\delta$ are. Then, the larger the sample-size $\gamma_T$ is, the less privacy noises are required to achieve the same $(\epsilon,\delta)$-differential privacy, and thus, the effect of privacy noises $d_{i,k}$ is reduced.
\end{rmk}
\begin{rmk}
	The sample-size $\gamma_T$ is not required to go to infinity to achieve finite cumulative differential privacy budgets $\epsilon$, $\delta$ over infinite iterations. Specifically, let the sample-size parameter $s=0$. Then, the sample-size $\gamma_T$ is constant. In this case, if $u-v>\max\{1-w,0\}$, $t\geq 2$, then Algorithm \ref{algorithm1} can achieve finite cumulative differential privacy budgets $\epsilon$, $\delta$ over infinite iterations. This result shows advantage over \cite{zhang2018improving,li2018differentially,huang2019dp,ding2021differentially,gratton2021privacy,xu2022dp,liu2024distributed,wang2023decentralized,yan2023killing} that only achieve infinite cumulative differential privacy budgets $\epsilon$, $\delta$ over infinite iterations and \cite{wang2023quantization} that only achieves $(0,\delta)$-differential privacy at each iteration.
\end{rmk}\vspace{-1.2em}
\subsection{Convergence analysis}
In this subsection, we will give the convergence analysis of Algorithm \ref{algorithm1}. First, we introduce an assumption on step-sizes, the sample-size and the privacy noise parameter.\vspace{-0.1em}
\begin{asm}\label{assumption 4: step sizes}
For any $T=0,1,\dots,k=0,\dots, T$, step-sizes $\alpha_T$$=$$\frac{a_1}{(T+1)^{u}}$,$\beta_T$$=$$\frac{a_2}{(T+1)^{v}}$, the sample-size $\gamma_T$$=$$\lfloor a_3 T^{s}\rfloor+1$ and the privacy noise parameter $\sigma_k=(k+1)^{w}$ satisfy $a_1$, $a_3>0$, $0<a_2<1$, $2u-v>1$, $\frac{1}{2}+\max\{w,0\}<v<u<1$.
\end{asm}

Next, we first provide the mean square convergence of Algorithm \ref{algorithm1}, and then show a convergence rate of Algorithm \ref{algorithm1} for cost functions satisfying the Polyak-{\L}ojasiewicz condition.

{\color{MidnightBlue}\small\sffamily\slshape \hspace{0.3cm}1) Mean square convergence}

Since saddle points make finding an optimal solution of the problem \eqref{question of interest 2} NP-hard (\!\!\cite{murty1987some}),  finding a first-order stationary point rather than an optimal solution is actually the main goal for distributed nonconvex stochastic optimization algorithms (see e.g. \cite{jiang2017collaborative,lu2023convergence,ge2023accelerate,ding2021differentially,xu2022dp,wang2023quantization,chen2024communication,reisizadeh2019robust}). Inspired by \cite{wang2023quantization},  $\E\|\nabla F(x_{i,T+1})\|^2$ is used as an index to show the  mean square convergence of Algorithm \ref{algorithm1}.
\begin{thm0}\label{thm2}
	If Assumptions \ref{assumption 1: graph preliminaries}-\ref{assumption 4: step sizes} hold, then\vspace{-0.5em}
	$$\liminf_{T\to\infty}\E\|\nabla F(x_{i,T+1})\|^2=0,  \forall i\in\mathcal{V}.$$
\end{thm0}
{\bf Proof.} See Appendix \ref{appendix1}. \hfill$\blacksquare$
\begin{rmk}
	In Theorem \ref{thm2}, by constructing an auxiliary variable $Y_k$$=$$\frac{1}{n}$$((I_n$$-$ $\frac{\mathbf{1}_n\mathbf{1}_n^{\!\top}}{n})$$\otimes$$I_r)x_k$, the convergence of Algorithm \ref{algorithm1} is achieved without assuming that gradients are bounded. This result shows advantage over \cite{zhang2023communication} that does not provide a convergence analysis, \cite{ge2023accelerate,jiang2017collaborative,xu2022dp,gratton2021privacy,yan2023killing,rabbat2005quantized,chen2024communication} that cannot achieve the mean square convergence, and \cite{lu2023convergence,li2018differentially,huang2019dp,ding2021differentially,wang2023quantization,wang2023decentralized,liu2024distributed} that assume the gradients are bounded. Thus, this new convergence technique has wider applicability than those in \cite{gratton2021privacy,rabbat2005quantized,chen2024communication,lu2023convergence,liu2024distributed,li2018differentially,huang2019dp,ding2021differentially,xu2022dp,wang2023decentralized,wang2023quantization,yan2023killing}.
\end{rmk}
{\color{MidnightBlue}\small\sffamily\slshape \hspace{0.3cm}2) Convergence rate analysis}
\begin{asm}\label{asm5}
	(Polyak-{\L}ojasiewicz) The global cost function $F(x)$ satisfies the Polyak-{\L}ojasiewicz condition, i.e., there exists $\mu>0$ such that $2\mu$$(F(x)-F^*)$$\leq$$\|\nabla F(x)\|^2$ for any $x$$\in$$\mathbb{R}^{r}$, where $F^*$ is the global minimum of the problem \eqref{question of interest 2}.
\end{asm}
\begin{rmk}\label{rmk9}
	Assumption \ref{asm5} is commonly used (see e.g. \cite{lu2023convergence}), and means that the gradient $\nabla F(x)$ to grow faster than a quadratic function as the algorithm moves away from the optimal solution. Such functions exist, for example, $F(x)=x^2+3\sin^2 x$ is a nonconvex function satisfying Assumption \ref{asm5} for any $ 0<\mu<0.3$. As shown in Theorem 2 of \cite{karimi2016linear}, Assumption \ref{asm5} is more general than the convex cost functions assumed in \cite{lei2022distributed,zhang2018improving,li2018differentially,huang2019dp,ding2021differentially,gratton2021privacy,liu2024distributed}.
	\end{rmk}
\begin{thm0}\label{thm3}
	If Assumptions \ref{assumption 1: graph preliminaries}-\ref{asm5} hold, then\vspace{-0.3em}
	$$\smash{\E\|\nabla F(x_{i,T\!+\!1})\|^\psi\!\!=\!\!O(\frac{\Delta^2}{(T\!+\!1)^{\frac{\psi}{2}\min\{2v-2\max\{w,0\}-1,2u-v-1\}}}),}$$
 for any $i\in\mathcal{V}$, $T=0,1,\dots$, and $\psi\in[1, 2]$. Particularly, when $\psi$$=$$2$, \mbox{we have}\vspace{-0.5em}
\begin{align}\label{eq21}
	\hspace{-0.9em}\smash{\E(F(x_{i,T\!+\!1})\!\!-\!\!F^*)\!\!=\!\!O(\frac{\Delta^2}{(T\!\!+\!\!1)^{\min\{2v\!-\!2\max\{w,0\}\!-\!1,2u\!-\!v\!-\!1\}}}),}
\end{align}{\vskip -2pt}\noindent where the constant in the big-$O$ notation does not depend on $\Delta$. Further, the mean square convergence of Algorithm~\ref{algorithm1} is achieved as $T$ goes to infinity, i.e.,  $\lim_{T\to\infty}\E\|\nabla F(x_{i,T+1})\|^2=0$, $\forall i\in\mathcal{V}$.
\end{thm0}
{\bf Proof.} See Appendix \ref{appendix2}. \hfill$\blacksquare$
\begin{rmk}
	To eliminate the effect of the quantization error on the convergence of Algorithm \ref{algorithm1}, a two-time-scale step-sizes method is used. The fast step-size $\alpha_T$ is used in the stochastic gradient descent, and the slow step-size $\beta_T$ is used to eliminate the effect of the quantization error on convergence. By Assumption \ref{assumption 4: step sizes}, the slow step-size $\beta_T$ satisfies $\lim_{T\to\infty}\beta_T^2\Delta^2=0$, which ensures the mean square convergence of Algorithm~\ref{algorithm1}. Compared with \cite{rabbat2005quantized,yan2023killing,chen2024communication}, the mean square convergence of Algorithm \ref{algorithm1} is achieved while improving communication efficiency simultaneously. Meanwhile, the problem of increasing network bandwidth in \cite{faghri2020adaptive} is solved. Moreover, \eqref{eq21} in Theorem \ref{thm3} shows the effect of the quantization error on the convergence rate, which is not considered in \cite{ding2021differentially,wang2023quantization}. The larger the quantization error $\Delta$ is, the slower the convergence rate is. Therefore, the probabilistic quantization does slow down the convergence rate of Algorithm \ref{algorithm1}.
\end{rmk}
\begin{rmk}
	The mean square convergence of Algorithm \ref{algorithm1} is guaranteed for general privacy noises, including increasing, constant (see e.g. \cite{li2018differentially,ding2021differentially,gratton2021privacy,liu2024distributed,xu2022dp,wang2023decentralized}) and decaying (see e.g. \cite{zhang2018improving,huang2019dp}) privacy noises. This is non-trivial even without considering privacy protection problem. For example, let $\alpha_T\!=\!\frac{1}{T^{0.9}},\beta_T\!=\!\frac{1}{T^{0.75}}$. Then, the convergence of Algorithm \ref{algorithm1} holds as long as the variance $\sigma_k$ of the privacy noise has an increasing rate no more than $O(k^{0.25})$.
\end{rmk}
\begin{rmk}
	Note that by Theorem \ref{thm2}, the mean square convergence of Algorithm \ref{algorithm1} holds for general cost functions, including convex and nonconvex cost functions. Then, when the global cost function is convex, Theorem \ref{thm2} also holds. Furthermore, if the global cost function $F(x)$ is $\lambda$-strongly convex, i.e., there exists $\lambda>0$ such that for any $x,y\in\mathbb{R}^r$, $F(y)\geq F(x)+\langle\nabla F(x),y-x\rangle+\frac{\lambda}{2}\|y-x\|^2$, then by \cite[Lemma 6.9]{bubeck2015convex} we have $2\lambda(F(x)-F^*)\leq\|\nabla F(x)\|^2$, which means the global cost function $F(x)$ satisfies Assumption \ref{asm5}. Thus, Algorithm \ref{algorithm1} achieves the same convergence rate as Theorem \ref{thm3}.
\end{rmk}
\begin{rmk}
	Note that distributed nonconvex stochastic optimization algorithms may converge to a saddle point instead of the desired global minimum. Then, the discussion of the avoidance of saddle points is necessary. Assumption \ref{asm5} implies that each stationary point $x^*$ of $F$ satisfying $\nabla F(x^*)=0$ is a global minimum of $F$, and thus, guarantees the avoidance of saddle points discussed in \cite{wang2023decentralized}. Furthermore, compared with \cite{wang2023decentralized}, Assumption \ref{asm5} helps us to give a convergence rate of Algorithm \ref{algorithm1}.
%	Furthermore, compared with \cite{lu2023convergence,ge2023accelerate,zhang2018improving,li2018differentially,huang2019dp,ding2021differentially,gratton2021privacy,wang2023quantization,wang2023decentralized,liu2024distributed}, Assumption \ref{asm5} helps us to give the convergence rate of Algorithm \ref{algorithm1} 
\end{rmk}

In practice, the time and number of running a distributed stochastic optimization algorithm are usually limited by various constraints, while selecting the best one from lots of running results is very time-consuming. To address this issue, the following low-probability convergence rate of Algorithm \ref{algorithm1} is given based on Theorem \ref{thm3}.
\begin{Cor}\label{cor2}
	Under Assumptions \ref{assumption 1: graph preliminaries}-\ref{asm5},  $$F(x_{i,T+1})-F^*=O(\frac{1}{(T+1)^{\min\{2v-2\max\{w,0\}-1,2u-v-1\}}})$$ with probability at least $1$$-$$\delta^*$, for any $i$$\in$$\mathcal{V}$,  $T$$=$$0,1,\dots$, and $\delta^*\in(0,1)$.
\end{Cor}
{\bf Proof.} By Theorem \ref{thm3}, there exists $A_1$$>$$0$ that does not depend on $\Delta$ such that $\E(F(x_{i,T+1})$$-$$F^*)$$\leq$$\frac{A_1\Delta^2}{(T+1)^{\min\{2v-2\max\{w,0\}-1,2u-v-1\}}}$. Let $a$$=$$\frac{A_1\Delta^2}{\delta^* (T+1)^{\min\{2v-2\max\{w,0\}-1,2u-v-1\}}}$ for any $\delta^*\in(0,1)$. Then, by Markov's inequality \cite[Th. 4.1.1]{chow2012probability} we have
\begin{align}\label{prob rate, 2}
	\smash{\p\left(F(x_{i,T+1})\!-\!F^*>a\right)\le\frac{\E(F(x_{i,T+1})\!-\!F^*)}{a}\le\delta^*.}
\end{align}
{\vskip -3pt}\noindent {\fontsize{9.5}{1}\selectfont Thus, by \eqref{prob rate, 2} we have $F(\!x_{i,T\!+\!1}\!)\!-\!F^*\!\!\leq\!\!\frac{A_1\Delta^2}{\delta^{*}\! (T\!+\!1)^{\!\min\!\{\!2\!v\!-\!2\!\max\{\!w\!,\!0\!\}\!-\!1,2\!u\!-\!v\!-\!1\!\}}}$} $=$$O(\!\frac{1}{(T\!+\!1)^{\min\{2v-2\max\{w,0\}-1,2u-v-1\}}}\!)$ with probability at least $1-\delta^*$. Therefore, this corollary \mbox{is proved.} \hfill$\blacksquare$
\begin{rmk}
Corollary \ref{cor2} guarantees the convergence of a single running result with probability at least $1-\delta^*$, and thus, avoids spending time on selecting the best one from lots of running results. Moreover, from Theorem \ref{cor2}, it follows that the low-probability convergence rate is affected by the failure probability $\delta^*$. The larger the failure probability $\delta^*$ is, the faster the low-probability convergence rate is.
\end{rmk}\vspace{-1em}
\subsection{Trade-off between privacy and utility}
Based on Theorems \ref{thm1}-\ref{thm3}, the mean square convergence of Algorithm \ref{algorithm1} as well as the differential privacy with finite cumulative differential privacy budgets $\epsilon$, $\delta$ over infinite iterations can be established simultaneously, which is given in the following corollary:
\begin{Cor}\label{cor3}
	For any $T=0,1,\dots,k=0,\dots, T$, let\vspace{-0.5em}
\begin{align*}
		&\alpha_T=\frac{a_1}{(T+1)^{u}},\beta_T=\frac{a_2}{(T+1)^{v}},\gamma_T=\lfloor a_3T^{s}\rfloor+1,\cr
		&\sigma_k=(k+1)^{w},\delta_k=\frac{1}{(k+2)^{t}},a_1,a_3>0, 0<a_2<1.
\end{align*}{\vskip -6pt}\indent If Assumptions \ref{assumption 1: graph preliminaries}-\ref{assumption 3: quantizer}, \ref{asm5} hold, and $t\geq 2$, $\frac{1}{2}+\max\{w,0\}<v$$<u$$<1$, $u$$+$$s-v>\max\{1-w,0\}$, $2u$$-$$v$$>1$, then Algorithm \ref{algorithm1} achieves the mean square convergence and finite cumulative differential privacy budgets $\epsilon$, $\delta$ over infinite iterations simultaneously as the sample-size $\gamma_T$ goes to infinity.
\end{Cor}
{\bf Proof.} By Theorems \ref{thm1}-\ref{thm3}, this corollary is proved. \hfill$\blacksquare$
\begin{rmk}
Corollary \ref{cor3} holds even when privacy noises have increasing variances. For example, when $u\!=\!0.98,v\!=\!0.8,w\!=\!0.2,s\!=\!0.7,t\!=\!2.5$, or $u\!=\!0.9,v\!=\!0.6,w\!=\!0.05,s\!=\!0.8,t\!=\!2$, the conditions of Corollary \ref{cor3} hold. In this case, the differential privacy with finite cumulative privacy budgets $\epsilon$, $\delta$ over infinite iterations as well as the mean square convergence can be established simultaneously.
\end{rmk}
\begin{rmk}
The result of Corollary \ref{cor3} does not contradict the trade-off between privacy and utility. In fact, to achieve differential privacy, Algorithm \ref{algorithm1} incurs a compromise on the utility. However, different from \cite{gratton2021privacy,yan2023killing} which compromise convergence accuracy to enable differential privacy, Algorithm~\ref{algorithm1} compromises the convergence rate and the sample-size (which are also utility metrics) instead. From Corollary~\ref{cor3}, it follows that the larger the privacy noise parameter $\sigma_{k}$ is, the slower the mean square convergence rate is. Besides, the sample-size $\gamma_T$ is required to go to infinity when the mean square convergence of Algorithm \ref{algorithm1} and finite cumulative privacy budgets $\epsilon$, $\delta$ over infinite iterations are considered simultaneously. The ability to retain convergence accuracy makes our approach suitable for accuracy-critical scenarios.
\end{rmk}\vspace{-1em}
\subsection{Oracle complexity}
Since the sampling parameter-controlled subsampling method is employed in Algorithm \ref{algorithm1}, the total number of data samples to obtain an optimal solution is an issue worthy of attention. To show this, we give the definitions of $\eta$-optimal solutions and the oracle complexity as follows:
\begin{def0}\label{def4:optimal-solution}
	($\eta$-optimal solution) Given $\eta>0$, $x_T=[x_{1,T}^\top,\dots,x_{n,T}^\top]^\top$ is an $\eta$-optimal solution if $\E|F(x_{i,T})-F^*|<\eta$, $\forall i\in\mathcal{V}$.
\end{def0}
\begin{def0}\label{def5:oracle complexity}
	Given $\eta>0$, the oracle complexity is the total number of data samples to obtain an $\eta$-optimal solution $\sum_{k=0}^{N(\eta)}\gamma_T$, where $N(\eta)=\min\{T\hspace{-0.35em}:\hspace{-0.15em}x_T\text{ is an }\eta$-optimal solution$\}$.
\end{def0}

Based on Theorem \ref{thm3}, Definitions \ref{def4:optimal-solution} and \ref{def5:oracle complexity}, the oracle complexity of Algorithm \ref{algorithm1} for obtaining an $\eta$-optimal solution is given as follows:
\begin{thm0}\label{thm4:complexity}
	Given $0<\eta<\frac{2}{5}$, let $u=1-\frac{\eta}{8}$, $v=\frac{2}{3}+\frac{7\eta}{12}$, $w=\eta$, $s=\eta$. Then, under Assumptions \ref{assumption 1: graph preliminaries}-\ref{assumption 3: quantizer} and \ref{asm5}, the oracle complexity of Algorithm \ref{algorithm1} is $O(\eta^{-\frac{6+6\eta}{2-5\eta}})$.
\end{thm0}

{\bf Proof.} For the given $\eta>0$, let the iteration maximum in Algorithm \ref{algorithm1} be $N(\eta)$. Then, we have $\gamma_T=\lfloor a_3N(\eta)^\eta\rfloor+1\leq a_3N(\eta)^\eta+1$. Note that by Theorem \ref{thm3}, there exists a constant $C>0$ that does not depend on $\Delta$ such that\vspace{-0.2em}
\begin{align}\label{eq: oracle complexity, 1}
	\smash{\E|F(x_{i,T\!+\!1})\!-\!F^*\!|\!=\!\E(F(x_{i,T\!+\!1})\!-\!F^*)\!\le\!\frac{C\Delta^2}{(T\!+\!1)^{\frac{1}{3}-\frac{5\eta}{6}}}.}
\end{align} 
{\vskip -1pt}\noindent Then, when $T\ge\lfloor(\frac{C\Delta^2}{\eta})^{\frac{6}{2-5\eta}}\rfloor$, \eqref{eq: oracle complexity, 1} can be rewritten as
\begin{align}\label{eq: oracle complexity, 2}
	\smash{\E|F(x_{i,T\!+\!1})\!-\!F^*\!|\!\leq\!\frac{C\Delta^2}{(T\!\!+\!\!1)^{\frac{1}{3}\!-\!\frac{5\eta}{6}}}\!\!<\!\!\frac{C\Delta^2}{(\frac{C\Delta^2}{\eta})^{(\frac{1}{3}\!-\!\frac{5\eta}{6})\frac{6}{2\!-\!5\eta}}}\!\!=\!\!\eta.}
\end{align}
{\vskip 2pt}\noindent Thus, by \eqref{eq: oracle complexity, 2} and Definition \ref{def4:optimal-solution}, $x_{T+1}$ is an $\eta$-optimal solution.

Since $N(\eta)$ is the smallest integer such that $x_{N(\eta)}$ is an $\eta$-optimal solution, we have\vspace{-0.9em}
\begin{align}\label{eq: oracle complexity, 3}
	N(\eta)\leq&1+\min\{T:T\geq\lfloor(\frac{C\Delta^2}{\eta}\!)^{\frac{6}{2-5\eta}}\rfloor\}\cr
	=&\smash{\lfloor(\frac{C\Delta^2}{\eta})^{\frac{6}{2-5\eta}}\rfloor+1.}
\end{align}
{\vskip -2pt}\indent Hence, by Definition \ref{def5:oracle complexity} and \eqref{eq: oracle complexity, 3}, we have\vspace{-0.7em} 
\begin{align*}
	\sum_{k=0}^{N(\eta)}\gamma_T=&(N(\eta)+1)\gamma_T\leq(N(\eta)+1)(a_3N(\eta)^\eta+1)\cr
	\noalign{\vskip -10pt}
	=&O\left(N(\eta)^{1+\eta}\right)=O\left(\eta^{-\frac{6+6\eta}{2-5\eta}}\right).
\end{align*}
{\vskip -5pt}\indent Therefore, this theorem is proved. \hfill$\blacksquare$
\begin{rmk}
From Theorems \ref{thm3} and \ref{thm4:complexity}, the faster the convergence rate is, the smaller the oracle complexity is. It is worth noting that as $T$ becomes large, one might question how one deals with $\gamma_T$ going to infinity. This issue does not arise in machine learning due to $\eta$-optimal solution is interested. For example, if $\eta=0.02$, then the total number of data samples to obtain an $\eta$-optimal solution is $O(10^5)$, which does not go to infinity. This requirement for the total number of data samples is acceptable since the computational cost of centralized stochastic gradient descent is $O(10^5)$ to achieve the same accuracy as Algorithm 1.
\end{rmk}\vspace{-0.3em}
\section{Numerical examples}\label{section4}
\begin{figure}[!htb]
	\centering
	\includegraphics[width=0.18\textwidth]{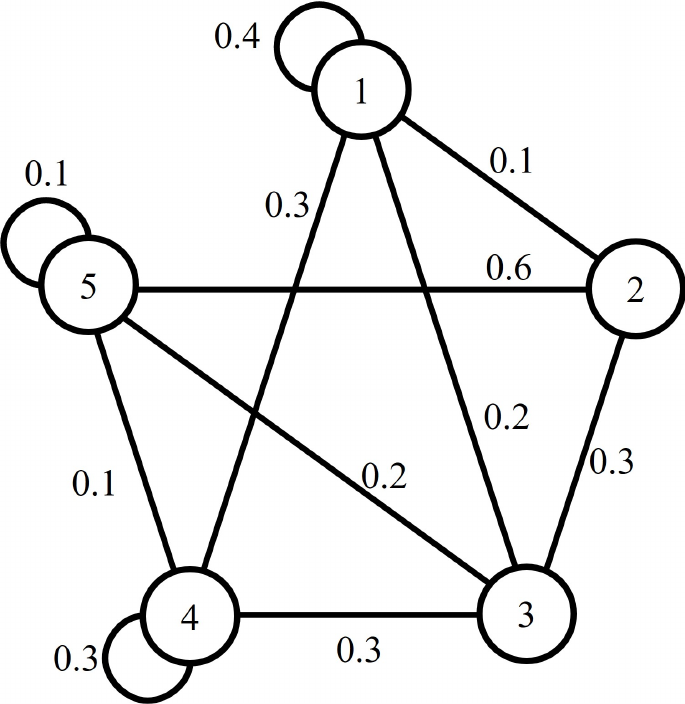}\vspace{-0.5em}
	\caption{Topology structure of the undirected graph}
	\label{fig1}\vspace{-1em}
\end{figure}

In this section, we verify the effectiveness and advantages of Algorithm 1 by the distributed training of a convolutional neural network (CNN) on the ``MNIST'' dataset (\!\!\cite{mnist1998lecun}). Specifically, five nodes cooperatively train a CNN using the ``MNIST'' dataset over a topology depicted in Fig. \ref{fig1}, which satisfies Assumption \ref{assumption 1: graph preliminaries}. Then, the ``MNIST'' dataset is divided into two subdatasets for training and testing, respectively. The training dataset is uniformly divided into 5 subdatasets consisting of 12000 binary images, and each of them can only be accessed by one agent to update its model parameters. The CNN model has two convolutional layers with 16, 32 filters, respectively, followed by a fully connected layer. The activation function of each convolutional layer is the Sigmoid function $\varphi(x)=\frac{1}{1+e^{-x}}$. Then, the global cost function is nonconvex and satisfies Assumption \ref{asm5}. In the following, the effect of the noise and the quantization on convergence, the differential privacy level, and the comparison with methods in \cite{li2018differentially,huang2019dp,ding2021differentially,gratton2021privacy,liu2024distributed,xu2022dp,wang2023decentralized} are presented for Algorithm~\ref{algorithm1}, respectively.\vspace{-0.7em}
\subsection{\fontsize{9.7}{10}\selectfont Effect of the noise and the quantization on convergence}
\begin{figure}[!htb]
	\centering
	\subfloat[Training accuracy]{\includegraphics[width=0.25\textwidth]{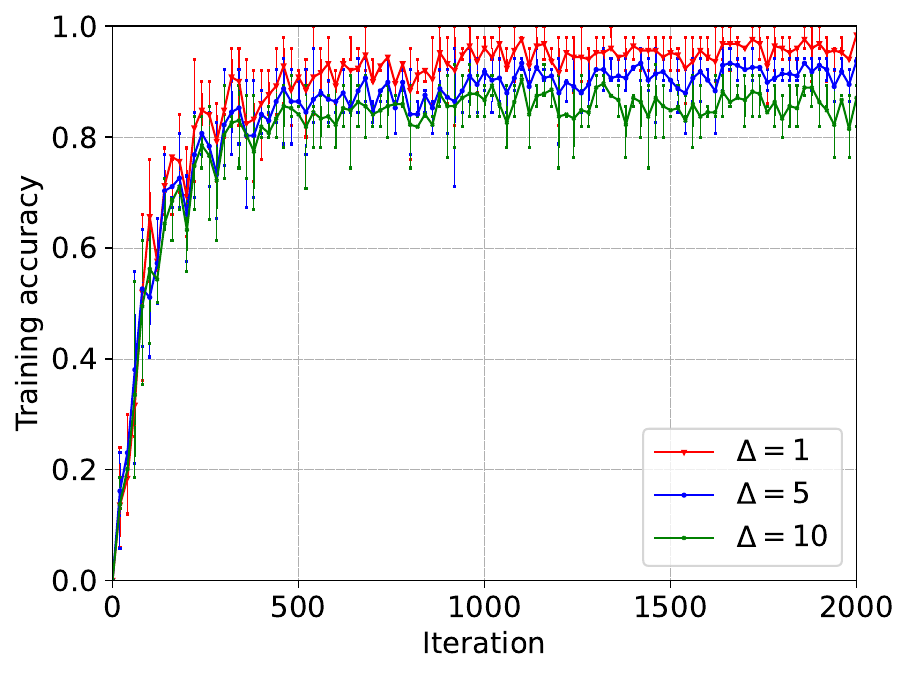}\label{fig2a}}
	\subfloat[Testing accuracy]{\includegraphics[width=0.25\textwidth]{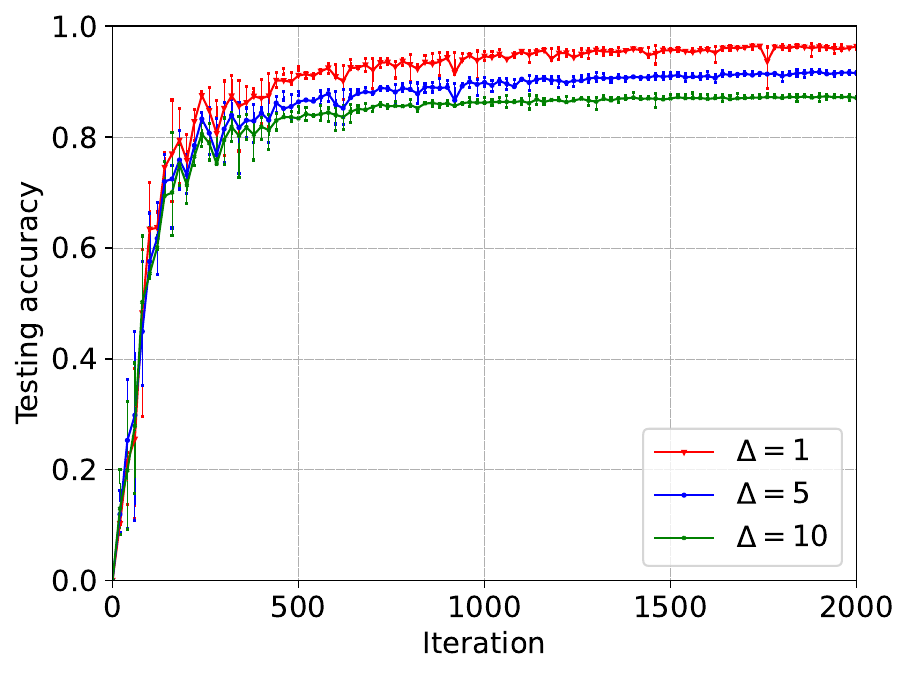}\label{fig2b}}\vspace{-0.3em}
	\caption{Accuracy of Algorithm \ref{algorithm1} with $\Delta =1,5,10$}
	\label{fig2}\vspace{-2em}
\end{figure}

\begin{figure}[!htb]
	\centering
	\subfloat[Training accuracy]{\includegraphics[width=0.25\textwidth]{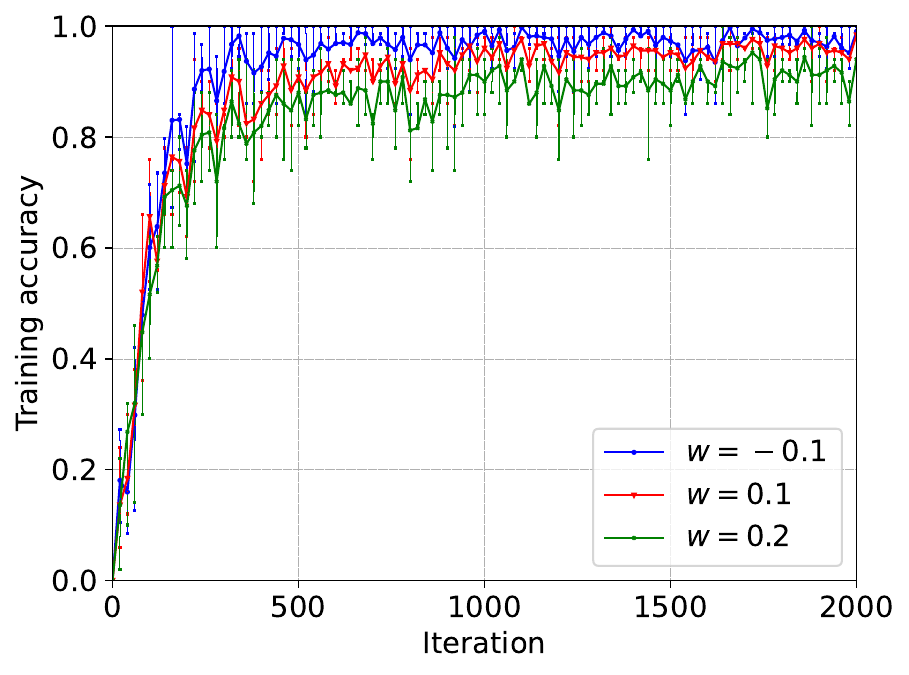}\label{fig3a}}
	\subfloat[Testing accuracy]{\includegraphics[width=0.25\textwidth]{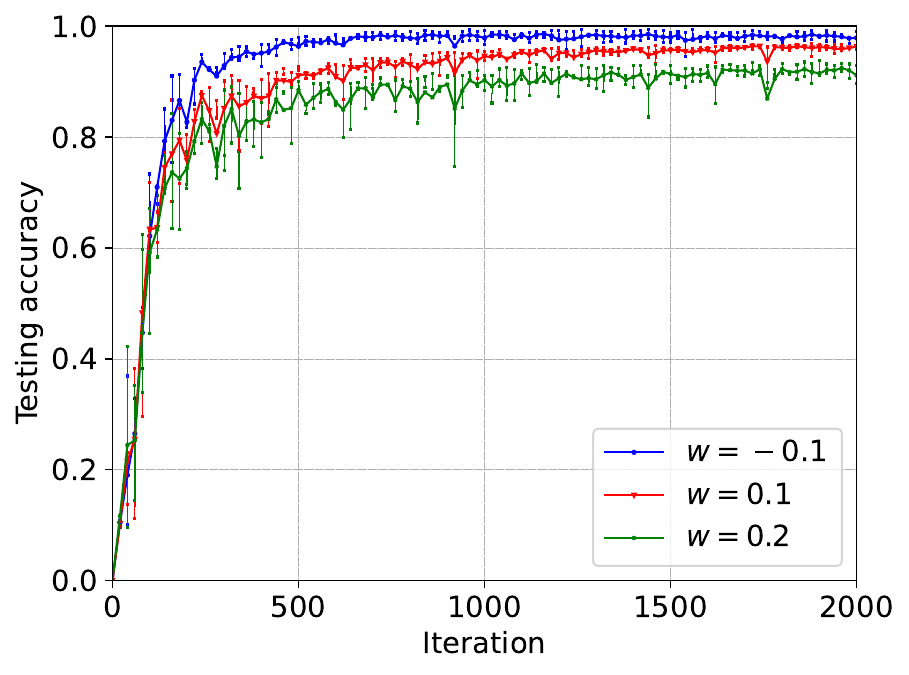}\label{fig3b}}\vspace{-0.2em}
	\caption{Accuracy of Algorithm \ref{algorithm1} with $w =-0.1,0.1,0.2$}
	\label{fig3}\vspace{-1.5em}
\end{figure}

Let step-sizes $\alpha_T\!=\!\frac{9.35}{2001^{0.9}}\approx10^{-2}$, $\beta_T\!=\!\frac{0.2}{2001^{0.7}}\approx10^{-3}$, the sample-size $\gamma_T\!=\!\lfloor5.5\cdot10^{-4}\cdot2000^{1.5}\rfloor+1=50$, $\delta_k=\frac{1}{(k+2)^3}$, and the privacy noise parameter $\sigma_k=(k+1)^{w}$ with $w=-0.1,0.1,0.2$, respectively. The probabilistic quantizer is given in the form of \eqref{quantizer mainly used} with $\Delta=1,5,10$, respectively. Then, it can be seen that Assumptions \ref{assumption 2: cost functions}-\ref{assumption 4: step sizes} hold. The training and testing accuracy on the ``MNIST'' dataset are presented in Figs. \ref{fig2} and \ref{fig3}, from which one can see that as iterations increase, the training and testing accuracy increase. More importantly, the smaller $\Delta$ and $w$ are, the faster Algorithm \ref{algorithm1} converges, which is consistent with Theorem \ref{thm3}.%\vspace{-1em}
\subsection{Differential privacy level}
\begin{figure}[!htb]
	\centering
	\includegraphics[width=0.4\textwidth]{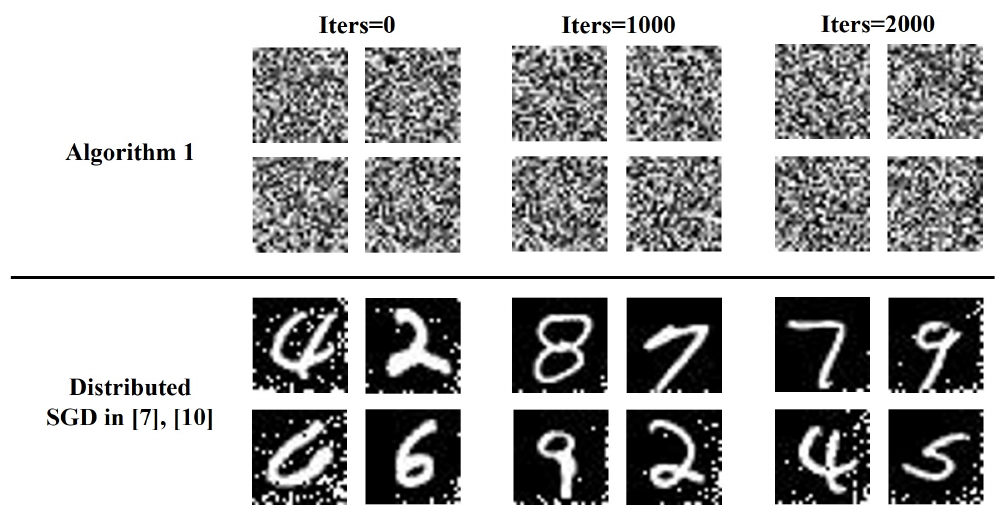}\vspace{-0.3em}
	\caption{ Comparison of privacy protection between \\Algorithm \ref{algorithm1} and distributed SGD in \cite{jiang2017collaborative,lu2023convergence}}
	\label{fig6}
\end{figure}

\vspace{-0.3em}
Based on the model inversion attack given in \cite{zhu2019deep}, we compare Algorithm \ref{algorithm1} and the algorithms without privacy protection in \cite{jiang2017collaborative,lu2023convergence} to show that Algorithm \ref{algorithm1} can protect the sensitive information from sampled gradients. A comparison of privacy protection between Algorithm 1 and distributed stochastic gradient descent (SGD) on the ``MNIST'' dataset is presented in Fig. \ref{fig6}, from which one can see that adversaries cannot recover original handwritten digit images in Algorithm \ref{algorithm1}, while adversaries can completely recover original handwritten digit images in distributed SGD (\!\!\cite{jiang2017collaborative,lu2023convergence}).

\begin{figure}[!htb]
	\centering
	\includegraphics[width=0.3\textwidth]{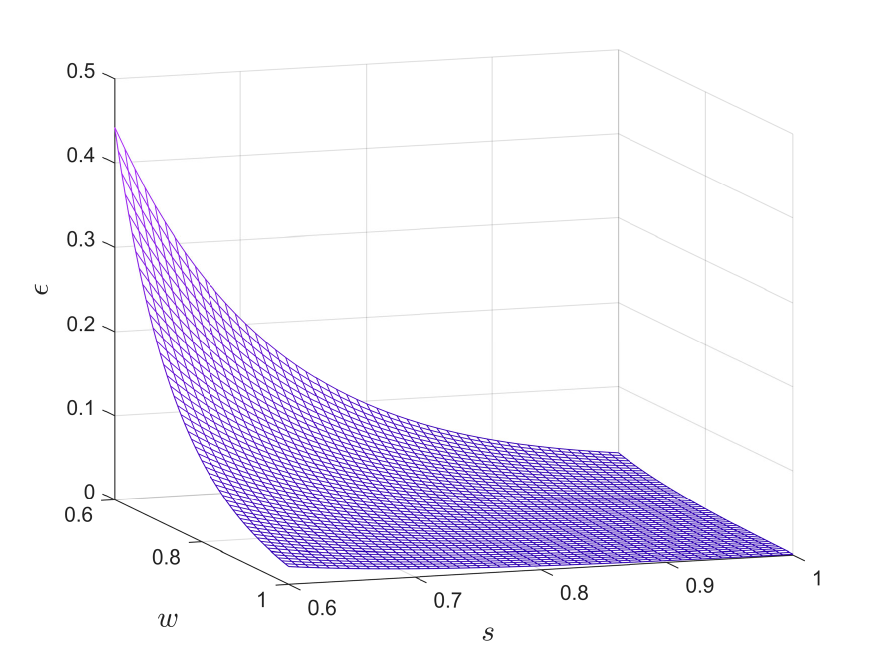}\vspace{-0.3em}
	\caption{Relationship of $\epsilon$, $w$ and $s$}
	\label{fig8}\vspace{-1.5em}
\end{figure}

Next, the relationship of the cumulative differential privacy budget $\epsilon$ over infinite iterations, the privacy noise parameter $w$ and sample-size parameter $s$ is presented in Fig. \ref{fig8}, from which one can see that as the privacy noise parameter $w$ and the sample-size parameter $s$ increase, the cumulative differential privacy budget $\epsilon$ decrease. This is consistent with the privacy analysis in Subsection \ref{subsection 3.2}. Moreover, in the first 2000 iterations, the cumulative differential privacy budgets $\epsilon=0.7594$ and  $\delta=0.2021$, which is consistent with Theorem~\ref{thm1}.\vspace{-0.5em}
\subsection{Comparison with methods in \cite{li2018differentially,huang2019dp,ding2021differentially,gratton2021privacy,xu2022dp,wang2023decentralized,liu2024distributed}}\label{4-c}
\begin{figure}[!htb]
	\centering
	\subfloat[Training accuracy]{\includegraphics[width=0.25\textwidth]{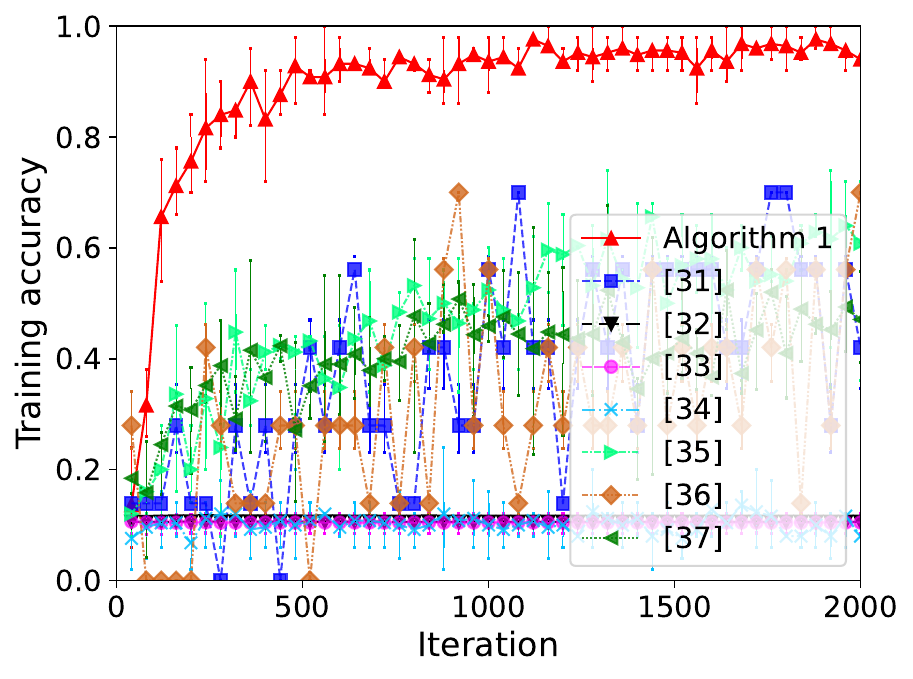}\label{fig4a}}
	\subfloat[Testing accuracy]{\includegraphics[width=0.25\textwidth]{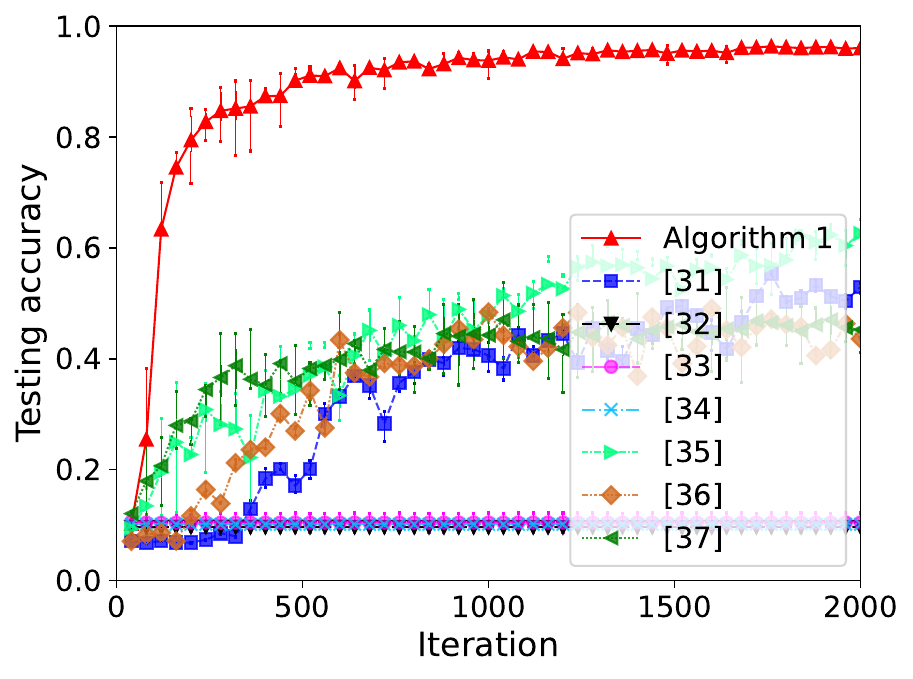}\label{fig4b}}\vspace{-0.4em}
	\caption{Comparison of accuracy}
	\label{fig4}\vspace{-3.5em}
\end{figure}

Let $\Delta=1,w=0.1$ in Algorithm \ref{algorithm1}. Then, the comparison of accuracy between Algorithm \ref{algorithm1} and methods in \cite{li2018differentially,huang2019dp,ding2021differentially,gratton2021privacy,xu2022dp,wang2023decentralized,liu2024distributed} is presented in Fig. \ref{fig4}. To ensure a fair comparison, we set the same step-sizes in \cite{li2018differentially,ding2021differentially,wang2023decentralized} as this paper, and the step-sizes in \cite{huang2019dp,gratton2021privacy,xu2022dp,liu2024distributed} as chosen therein. In addition, we set sample-sizes in \cite{li2018differentially,huang2019dp,ding2021differentially,gratton2021privacy,xu2022dp,wang2023decentralized,liu2024distributed} as chosen therein. From Figs. \ref{fig4}\subref{fig4a} and \ref{fig4}\subref{fig4b}, it can be seen that the convergence rate of Algorithm~\ref{algorithm1} is faster than \cite{li2018differentially,huang2019dp,ding2021differentially,gratton2021privacy,xu2022dp,wang2023decentralized,liu2024distributed}.

\begin{figure}[!htb]
	\centering
	\subfloat[$\epsilon$ under same~$\sigma_k,\delta_k$]{\includegraphics[width=0.255\textwidth]{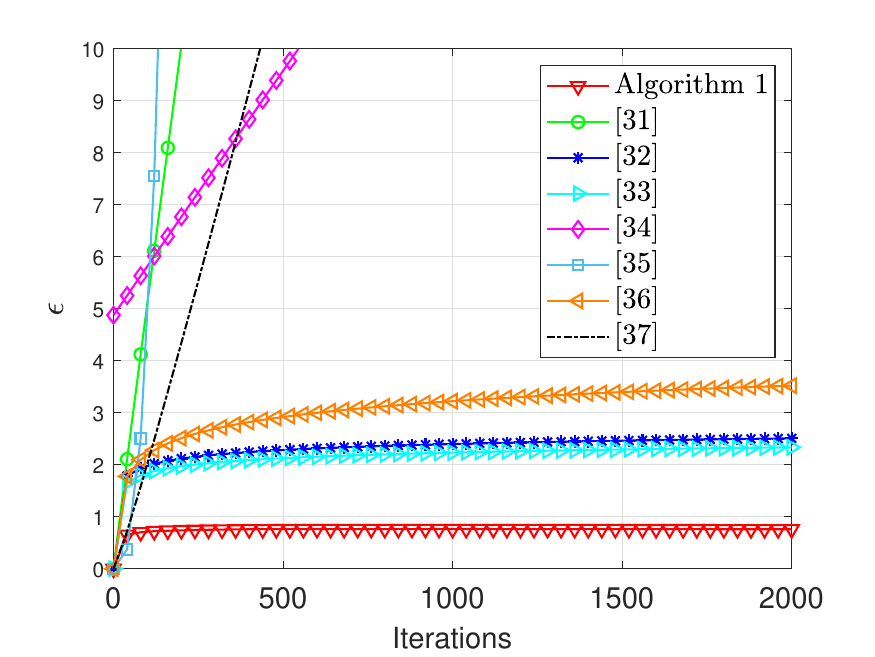}\label{fig7a}}
	\subfloat[$\delta$ under same~$\sigma_k,\epsilon_k$]{\includegraphics[width=0.255\textwidth]{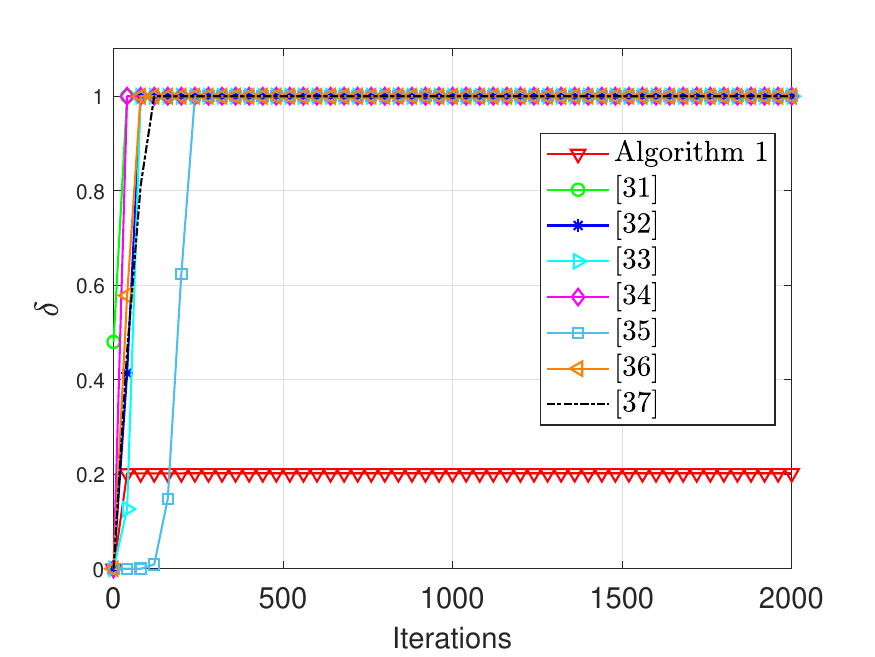}\label{fig7b}}\vspace{-0.3em}
	\caption{Comparison of cumulative differential privacy budgets $\epsilon$ and $\delta$}
	\label{fig7}\vspace{-2em}
\end{figure}

Since the structure of the CNN model is known, the sampled gradient $\|\nabla \ell_i(x,\xi_{i,l})\|$ is bounded for any $x\in\mathbb{R}^{29034}$ and $\xi_{i,l}\in \mathcal{D}$. When running the CNN model on the ``MNIST'' dataset, the maximum magnitude of the sampled gradient $\|\nabla \ell_i(x,\xi_{i_0,l_0})-\nabla \ell_i(x,\xi_{i_0,l_0}^\prime)\|$ is no more than 60 after changing one data sample $\xi_{i_0,l_0}$ to any different data sample $\xi_{i_0,l_0}^\prime$. Then, when the constant $C=60$, Definition \ref{def1:adjacency relation} contains the adjacency relation in \cite{li2018differentially,huang2019dp,ding2021differentially,gratton2021privacy,xu2022dp,wang2023decentralized,liu2024distributed} and vice versa, which implies that Definition 1 is equivalent to the adjacency relation therein. Thus, cumulative differential privacy budgets $\epsilon,\delta$ of Algorithm \ref{algorithm1} can be compared with those of methods in \cite{li2018differentially,huang2019dp,ding2021differentially,gratton2021privacy,xu2022dp,wang2023decentralized,liu2024distributed}, and the comparison of cumulative differential privacy budgets $\epsilon,\delta$ is presented in Fig. \ref{fig7}. From Figs. \ref{fig7}\subref{fig7a} and \ref{fig7}\subref{fig7b} one can see that cumulative differential privacy budgets $\epsilon,\delta$ of Algorithm \ref{algorithm1} are bounded by finite constants over infinite iterations, while cumulative differential privacy budgets $\epsilon,\delta$ in \cite{li2018differentially,huang2019dp,ding2021differentially,gratton2021privacy,xu2022dp,wang2023decentralized,liu2024distributed} go to infinity over infinite iterations.

\begin{figure}[!htb]
	\centering
	\subfloat[The ``MNIST'' dataset, $C=60$]{\includegraphics[width=0.47\textwidth]{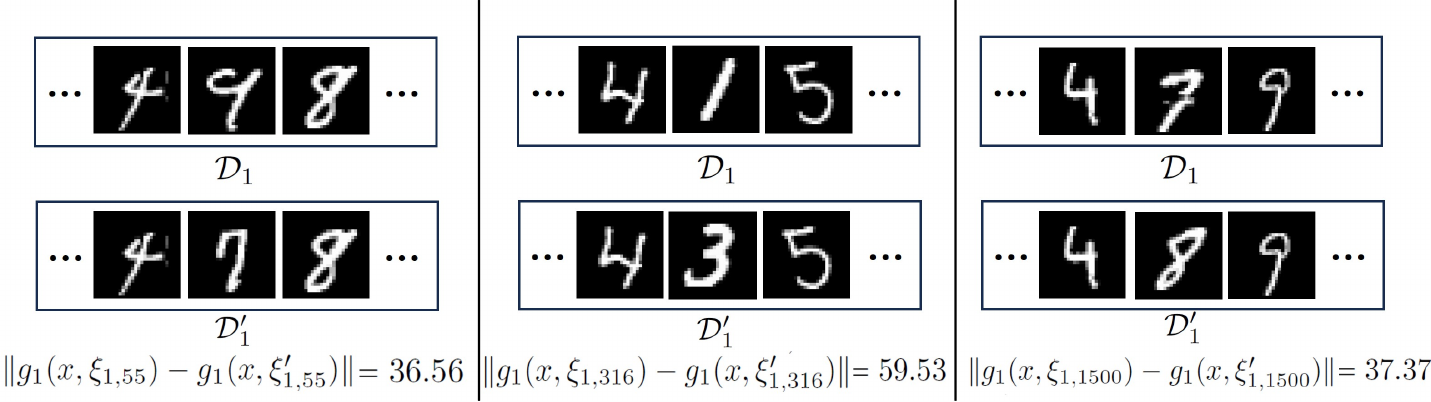}\label{fig5a}}\vspace{-0.9em}\\
	\subfloat[The ``CIFAR-10'' dataset, $C=20$]{\includegraphics[width=0.47\textwidth]{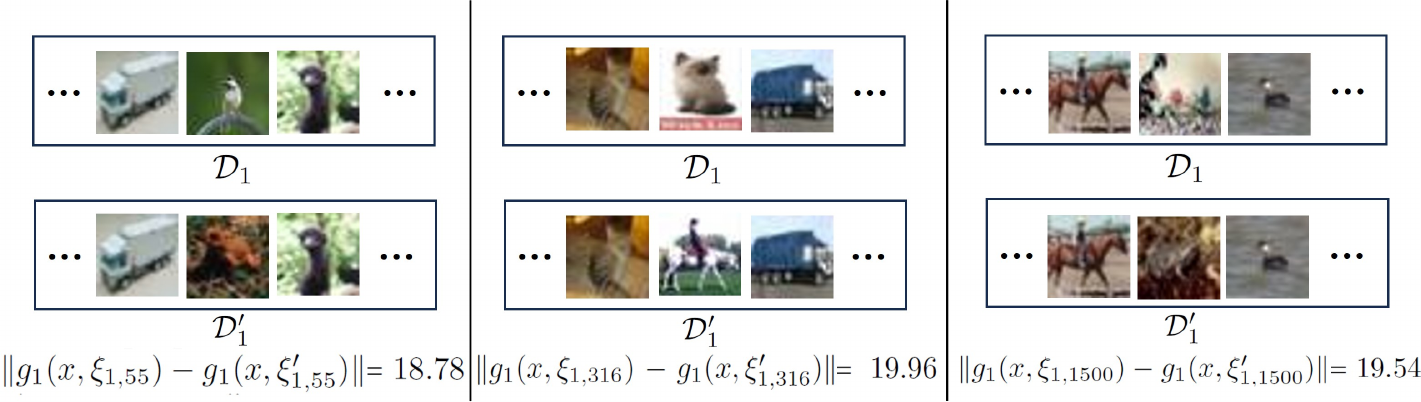}\label{fig5b}}\vspace{-0.9em}\\
	\subfloat[The ``CIFAR-100'' dataset, $C=19.5$]{\includegraphics[width=0.47\textwidth]{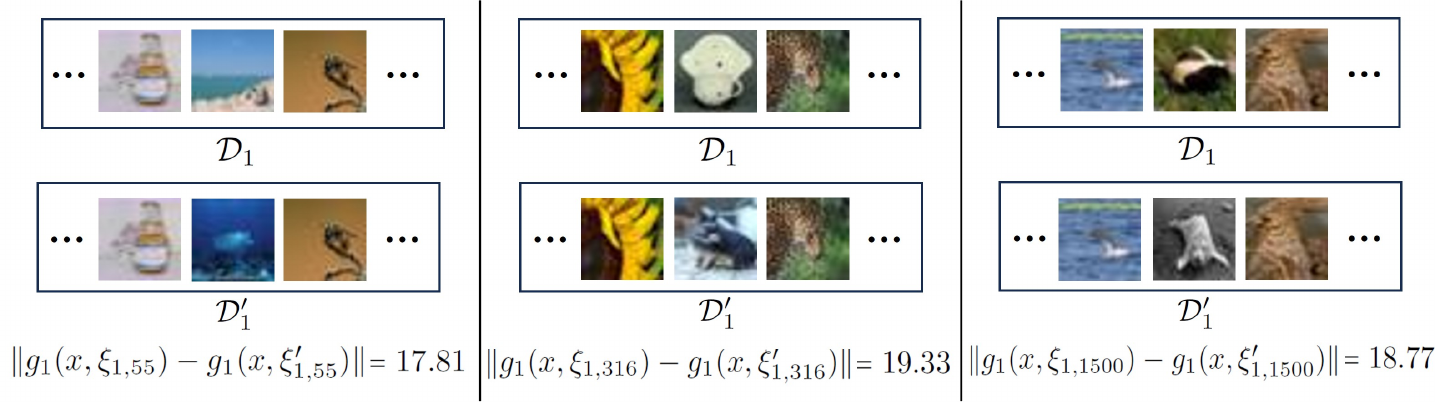}\label{fig5c}}\vspace{-0.4em}
	\caption{ Different constant $C$ for different datasets}
	\label{fig5}\vspace{-2em}
\end{figure}

In summary, the discussion above demonstrates Algorithm~\ref{algorithm1}'s superior performance over \cite{li2018differentially,huang2019dp,ding2021differentially,gratton2021privacy,xu2022dp,wang2023decentralized,liu2024distributed} on the convergence rate and the differential privacy level.
%Based on the above discussions, Algorithm \ref{algorithm1} not only converges, but also provides smaller cumulative differential privacy budgets $\epsilon$, $\delta$ over infinite iterations.
\begin{rmk}
	It is noted that only when a comparison between the method of this paper and the methods in \cite{li2018differentially,huang2019dp,ding2021differentially,gratton2021privacy,xu2022dp,wang2023decentralized,liu2024distributed} is needed, the constant $C$ can be different for different datasets. For example, Fig.~\ref{fig5} shows different constant $C$ for the ``MNIST'', ``CIFAR-10''\cite{cifar10torento} and ``CIFAR-100''\cite{cifar100torento,cifar} dataset, respectively. For each dataset, we randomly change one data sample and compute the magnitude of sampled gradients. Due to the space limitation, only three examples are given for each dataset. Fig. \ref{fig5}\subref{fig5a} shows that for the ``MNIST'' dataset, the magnitude of sampled gradients after respectively changing the 55th, 316th, 1500th data sample is 36.56, 59.53, 37.37, which is no more than the constant $C=60$. Similarly, Figs. \ref{fig5}\subref{fig5b} and \ref{fig5}\subref{fig5c} show that the magnitude of sampled gradients is no more than the constant $C=20$ and $19.5$, respectively. This interesting finding is consistent with \cite{li2018differentially,huang2019dp,ding2021differentially,gratton2021privacy,xu2022dp,wang2023decentralized,liu2024distributed}, where the upper bound of bounded gradients is also different for different datasets. When a comparison is not needed, the constant $C$ can be a fixed value for different datasets.
%	
%	This interesting finding is consistent with \cite{li2018differentially,huang2019dp,ding2021differentially,gratton2021privacy,xu2022dp,wang2023decentralized,liu2024distributed}, where the upper bound of bounded gradients is also different for different datasets.
\end{rmk}\vspace{-2em}
\section{Conclusion}\label{section5}
In this paper, we have proposed a differentially private distributed nonconvex stochastic optimization algorithm with quantized communication. In the proposed algorithm, general privacy noises are added to each node's local states to protect the sensitive information, and then a probabilistic quantizer is employed on noise-perturbed states to improve communication efficiency. By using the sampling parameter-controlled subsampling method, the differential privacy level of the algorithm is enhanced compared with the existing ones. By using a new convergence analysis technique and the two-time-scale step-sizes method, the effect of the quantization error on convergence is eliminated while improving communication efficiency, and thus, the mean square convergence for nonconvex cost functions is obtained. Then, under the Polyak-{\L}ojasiewicz condition, the mean square convergence rate and the oracle complexity of the algorithm are given. Meanwhile, the trade-off between the privacy and the utility is shown. Finally, a numerical example of the distributed training of CNN on the ``MNIST'' dataset is given to verify the effectiveness of \mbox{the algorithm.}

\vspace{-0.8em}
%{\vskip -5pt}
\appendices
\setcounter{lemma}{0}
\renewcommand{\thelemma}{A.\arabic{lemma}}
%{\vskip -10pt}
\section{ A useful lemma}\vspace{-0.3em}
%{\vskip -15pt}
\begin{lemma}\label{lemma a1}
If  a function $h(x)$ defined on $\mathbb{R}^{r}$ satisfies Assumption \ref{assumption 2: cost functions}(i), and $\min_{x\in\mathbb{R}^{r}}h(x)=h^*>-\infty$, then the following results hold: (i) $h(y)\leq h(x)+\langle \nabla h(x),y-x \rangle+\frac{L_1}{2}\|y-x\|^2$, $\forall x,y\in \mathbb{R}^{r}$; (ii) $\|\nabla h(x)\|^2\leq 2L_1\left(h(x)-h^*\right)$, $\forall x\in\mathbb{R}^{r}$.
\end{lemma}
{\bf Proof.} Lemma \ref{lemma a1}(i) is directly from \cite[Lemma 3.4]{bubeck2015convex}. By (3.5) in \cite{bubeck2015convex}, we have $\|\nabla h(x)\|^2$$\leq$$2L_1(h(x)$$-$ $h(x-$$\frac{1}{L_1}\nabla h(x)))$ $\leq$$2L_1(h(x)-h^*)$, then Lemma \ref{lemma a1}(ii) is proved. \hfill$\blacksquare$\vspace{-0.6em}

\section{Proof of Theorem \ref{thm2}}\label{appendix1}\vspace{-0.4em}
To provide an explanation of our results clearly, define\vspace{-0.5em}
\begin{align*}
	&\nabla f(x_k)\triangleq[{\nabla f_1(x_{1,k})}^\top,{\nabla f_2(x_{2,k})}^\top,\dots,{\nabla f_n(x_{n,k})}^\top]^\top,\cr
	&\nabla\ell(x_k)\triangleq[\nabla\ell_{1,k}^\top,\nabla\ell_{2,k}^\top,\dots,\nabla\ell_{n,k}^\top]^\top,\cr
	&\nabla f(\bar{x}_k)\triangleq[\nabla f_1(\bar{x}_k)^\top,\nabla f_2(\bar{x}_k)^\top,\dots,\nabla f_n(\bar{x}_k)^\top]^\top,\cr
	\noalign{\vskip 2pt}
	&\smash{W\triangleq I_n-\frac{1}{n}\mathbf{1}_{n}\mathbf{1}_n^\top,\; Y_k\triangleq(W\otimes I_{r})x_k,}\cr
	&e_k\triangleq z_k-x_k-d_k,\;w_k\triangleq \nabla\ell(x_k)-\nabla f(x_k),\cr
	\noalign{\vskip 2pt}
	&\smash{\bar{x}_k\triangleq\frac{1}{n}(\mathbf{1}_n^\top \otimes I_{r})x_k,\;\bar{w}_k\triangleq\frac{1}{n}(\mathbf{1}_n^\top \otimes I_{r})w_k,}\cr
	\noalign{\vskip 4pt}
	&\smash{\overline{\nabla f(x_k)}\triangleq\frac{1}{n}(\mathbf{1}_n^\top \otimes I_{r})\nabla f(x_k)=\frac{1}{n}\sum_{i=1}^{n}\nabla f_{i}(x_{i,k}).}
\end{align*}{\vskip 2pt}\indent Then, we can express \eqref{algorithm: per-node} for all nodes in a compact form as follows:\vspace{-1em}
\begin{align}\label{eq:compact form}
	x_{k+1}=&((I_n-\beta_T\mathcal{L})\otimes I_{r})x_k-\alpha_T \nabla f(x_k)\cr
	&+\beta_T (\mathcal{A}\otimes I_{r})(e_k+d_k)-\alpha_T w_k.
\end{align}
{\vskip -6pt}\noindent Next, the following six steps are given to prove Theorem \ref{thm2}.

\textbf{Step 1:} We first consider the term $\|Y_k\|^2$. Note that $W(I_n-\beta_T\mathcal{L})=(I_n-\beta_T\mathcal{L})W$. Then, multiplying both sides of \eqref{eq:compact form} by $W\otimes I_{r}$ gives\vspace{-0.5em}
\begin{align}\label{eq:vectorized consensus part}
	Y_{k+1}=&\left((I_n-\beta_T\mathcal{L})\otimes I_{r}\right)Y_k-\alpha_T(W\otimes I_{r})\nabla f(x_k)\cr
	&\!+\!\beta_T(W\mathcal{A}\!\otimes\!I_{r})(e_k\!+\!d_k)\!-\!\alpha_T(W\!\otimes\!I_{r})w_k.
\end{align}

{\vskip -4pt}\indent For any $k$$=$$0,$$\dots$$,$$ T$, define $\sigma$-algebras $\mathcal{F}_k$$=$$\sigma(x_k,d_k)$, $\mathcal{H}_k$$=$$\sigma(x_k)$. Then, since $d_{i,k}$ is independent of $\mathcal{H}_k$ and follows the normal distribution $N(0,\sigma_k^2 I_{r})$, we have \vspace{-0.7em}
\begin{align}
	&\E d_k=\E(d_k|\mathcal{H}_k)=0, \label{eq:expectation of nk}\\
	&\E\|d_k\|^2=\E(\|d_k\|^2|\mathcal{H}_k)= nr\sigma_k^2.\label{eq:variance of nk}
\end{align}{\vskip -7pt}

Since $w_k$ is independent of $\mathcal{F}_k$, by Assumption \ref{assumption 2: cost functions}(iii) we have\vspace{-1em}
\begin{align}
	&\E w_k=\E(w_k|\mathcal{F}_k)=0,\label{eq:expectation of wk}\\
	\noalign{\vskip -7pt}
	&\E\|w_k\|^2=\E(\|w_k\|^2|\mathcal{F}_k)\leq \frac{n\sigma_{\ell}^2}{\gamma_T}.\label{eq:variance of wk}
\end{align}{\vskip -7pt}
Since $e_k$ is independent of $\mathcal{F}_k$, by Assumption \ref{assumption 3: quantizer} we have \vspace{-0.7em}
\begin{align}
	&\E e_k=\E(e_k|\mathcal{F}_k)=0,\label{eq:expectation of ek}\\
	&\E\|e_k\|^2=\E(\|e_k\|^2|\mathcal{F}_k)\leq nr\Delta^2. \label{eq:variance of ek}
\end{align}{\vskip -7pt}

By \eqref{eq:expectation of nk}, \eqref{eq:expectation of wk} and \eqref{eq:expectation of ek}, taking mathematical expectation of $\|Y_{k+1}\|^2$ leads to
\begin{align}\label{eq: consensus part, 1}
	&\E\left\|Y_{k+1}\right\|^2\cr
	=&\E\!\left\|\!((I_n\!-\!\!\beta_T\mathcal{L})\!\otimes\! I_r\!)Y_k\!\!-\!\!\alpha_T(W\!\otimes\!I_r)\nabla f(x_k)\right.\cr
	&+\left.\!\beta_T(W\mathcal{A}\!\!\otimes\!\!I_r)(\!d_k\!\!+\!\!e_k\!)\!-\!\alpha_T(W\!\otimes\!I_r)w_k \right\|^2\cr
	=&\E\!\left\|\!((\!I_n\!\!-\!\!\beta_T\mathcal{L})\!\otimes\!I_r\!)Y_k\!\!-\!\!\alpha_T(W\!\otimes\!I_r)\nabla\!f(x_k)\right\|^2\cr
	&+\!\!\beta_T^2\E\left\|(W\mathcal{A}\!\otimes\!I_r)(\!d_k\!\!+\!\!e_k\!)\right\|^2\!+\!\!\alpha_T^2\E\!\left\|(W\!\!\otimes\!\!I_r)w_k\right\|^2\cr
	&\!\!+\!\!2\beta_T\E\!\left\langle\!\left((\!I_n\!\!-\!\!\beta_T\mathcal{L})\!\!\otimes\!\! I_r\!\right)\!\!Y_k\!\!-\!\!\alpha_T(W\!\!\otimes\!\!I_r)\!\nabla\!f(x_k)\!,(\!\mathcal{A}W\!\!\otimes\!\! I_r\!)(\!d_k\!\!+\!\!e_k\!)\right\rangle\nonumber\\
	&\!\!+\!\!2\alpha_T\E\!\left\langle\left((\!I_n\!\!-\!\!\beta_T\mathcal{L})\!\!\otimes\!\! I_r\!\right)\!\!Y_k\!\!-\!\!\alpha_T(W\!\!\otimes\!\!I_r)\!\nabla\!f(x_k)\!,(W\!\!\otimes\!\!I_r)w_k\!\right\rangle\cr
	&\!\!+\!\!2\alpha_T\beta_T\E\left\langle (W\mathcal{A}\otimes I_r)(d_k\!+\!e_k),(W\otimes I_r)w_k \right\rangle\cr
	=&\E\left\|\left((I_n\!-\!\beta_T\mathcal{L})\otimes I_r\right)Y_k\!-\!\alpha_T(W\otimes I_r)\nabla f(x_k)\right\|^2\cr
	&\!\!+\!\!\beta_T^2\E\left(\left\|(W\mathcal{A}\otimes I_r)(d_k\!+\!e_k)\right\|^2\right)\cr
	&\!\!+\!\!\alpha_T^2\E\left\|(W\otimes I_r)w_k\right\|^2.
\end{align}
Then, by the law of total expectation \cite[Th. 7.1.1]{chow2012probability}, we have\vspace{-0.5em}
\begin{align}\label{eq: consensus part, 2}
	&\E\langle\!(\!(\!W\mathcal{A}\!)\!\otimes\!I_r\!)d_k, (\!(\!W\mathcal{A}\!)\!\otimes\!I_r\!)e_k\rangle\!\cr
	=&\E(\E(\!\langle\!(\!(\!W\mathcal{A}\!)\!\otimes\!I_r\!)d_k, (\!(\!W\mathcal{A}\!)\!\otimes\!I_r\!)e_k\rangle|\mathcal{F}_k\!)\!)\cr
	=&\E\!\left\langle(\!(\!W\mathcal{A}\!)\!\otimes\!I_r\!)d_k,\E((\!(\!W\mathcal{A}\!)\!\otimes\!I_r\!)e_k|\mathcal{F}_k)\right\rangle\cr
	=&\E\left\langle (\!(\!W\mathcal{A}\!)\!\otimes\!I_r\!)d_k,0\right\rangle=0.
\end{align}{\vskip -4pt}\noindent Thus, substituting equation \eqref{eq: consensus part, 2} into equation \eqref{eq: consensus part, 1} implies
\begin{align}\label{eq: consensus part, 4}
	&\E\left\|Y_{k+1}\right\|^2\cr
	=&\E\left\|\left((I_n\!-\!\beta_T\mathcal{L})\otimes I_r\right)Y_k\!-\!\alpha_T(W\otimes I_r)\nabla f(x_k)\right\|^2\cr
	&+\beta_T^2\E\left(\left\|(W\mathcal{A}\otimes I_r)d_k\right\|^2\!+\!\left\|(W\mathcal{A}\otimes I_r)e_k\right\|^2\!\right)\cr
	&+2\E\!\left\langle(\!(\!W\mathcal{A}\!)\!\otimes\!I_r\!)d_k,(\!(\!W\mathcal{A}\!)\!\otimes\!I_r\!)e_k\! \right\rangle\!\!+\!\!\alpha_T^2\E\!\left\|(W\!\otimes\!I_r)w_k\right\|^2\cr
	=&\E\left\|\left((I_n\!-\!\beta_T\mathcal{L})\otimes I_r\right)Y_k\!-\!\alpha_T(W\otimes I_r)\nabla f(x_k)\right\|^2\cr
	&+\beta_T^2\E\left(\left\|(W\mathcal{A}\otimes I_r)d_k\right\|^2\!+\!\left\|(W\mathcal{A}\otimes I_r)e_k\right\|^2\right)\cr
	&+\alpha_T^2\E\left\|(W\otimes I_r)w_k\right\|^2.
\end{align}

By Rayleigh Theorem \cite[Th. 4.2.2]{horn2012matrix} and Assumption \ref{assumption 1: graph preliminaries}, $\|\mathcal{A}\|=1$. Note that $\|A\mathbf{x}\|\le\|A\|\|\mathbf{x}\|$ for any $A\in\mathbb{R}^{n\times n}$, $\mathbf{x}\in\mathbb{R}^{n}$. Then, by $\|W\|=1$, substituting \eqref{eq:variance of nk}, \eqref{eq:variance of wk} and \eqref{eq:variance of ek} into \eqref{eq: consensus part, 4}~implies\vspace{-0.5em}
\begin{align}\label{eq: consensus part, 3}
	\E\left\|Y_{k+1}\right\|^2\leq&\E\left\|\left(\!(I_n\!-\!\beta_T\mathcal{L})\!\otimes\!I_{r}\!\right)\!Y_k\!-\!\alpha_T(W\!\otimes\!I_{r})\!\nabla f(x_k)\right\|^2\nonumber\\
	&+nr\beta_T^2(\Delta^2\!+\!\sigma_k^2)\!+\!\frac{n\alpha_T^2\sigma_{\ell}^2}{\gamma_T}.
\end{align}
{\vskip -4pt}\noindent Furthermore, for any $\mathbf{a},\mathbf{b}\in\mathbb{R}^{r}$, the following Cauchy-Schwarz inequality \cite[Ex. 4(b)]{zorich2015analysis} holds: $
\|\mathbf{a}+\mathbf{b}\|^2\le(1+\rho_{\mathcal{L}}\beta_T)\|\mathbf{a}\|^2+(1+\frac{1}{\rho_{\mathcal{L}}\beta_T})\|\mathbf{b}\|^2$, where $\rho_{\mathcal{L}}>0$ is the second smallest eigenvalue of $\mathcal{L}$.  This together with \eqref{eq: consensus part, 3} gives
\begin{align}\label{eq:consensus ,1}
	\E\|Y_{k+1}\|^2\le&\left(1\!+\!\rho_{\mathcal{L}}\beta_T\right)\E\left\|\left(\left(\!I_n\!-\beta_T\mathcal{L}\right)\!\otimes\!I_r\!\right)\!Y_k\right\|^2\cr
	&+\left(1\!+\!\frac{1}{\rho_{\mathcal{L}}\beta_T}\right)\E\|\alpha_T(W\!\otimes\!I_{r})\nabla f(x_k)\|^2\cr
	&+\frac{n\alpha_T^2\sigma_{\ell}^2}{\gamma_T}+n{r}\beta_T^2(\Delta^2+\sigma_k^2).
\end{align}

\vspace{-0.2em}By Courant-Fischer's Theorem  \cite[Th. 4.2.6]{horn2012matrix} we have
\begin{align}\label{eq:consensus ,1.1}
	\smash{\left\|\left(\left(\!I_n\!-\beta_T\mathcal{L}\right)\!\otimes\!I_{r}\!\right)\!Y_k\right\|^2\le(1-\rho_{\mathcal{L}}\beta_T)^2\|Y_k\|^2.}
\end{align}
Thus, substituting \eqref{eq:consensus ,1.1} into \eqref{eq:consensus ,1} and noticing $\|W\|=1$, one can get\vspace{-0.4em}
\begin{align}\label{eq:consensus ,1.5}
	&\E\|Y_{k+1}\|^2~~~~~~~~~~~~~~~~~~~~~~~~\cr
	\leq&(1+\rho_{\mathcal{L}}\beta_T)(1\!-\!\rho_{\mathcal{L}}\beta_T)^2\E\|Y_k\|^2+nr\beta_T^2(\Delta^2\!+\!\sigma_k^2)\cr
	&+\frac{1+\rho_{\mathcal{L}}\beta_T}{\rho_{\mathcal{L}}\beta_T}\E\left\|\alpha_T(W\!\otimes\!I_{r})\nabla f(x_k)\right\|^2+\frac{n\alpha_T^2\sigma_{\ell}^2}{\gamma_T}\cr
	\leq&(1+\rho_{\mathcal{L}}\beta_T)(1\!-\!\rho_{\mathcal{L}}\beta_T)^2\E\|Y_k\|^2+nr\beta_T^2(\Delta^2\!+\!\sigma_k^2)\cr
	&+\frac{(1+\rho_{\mathcal{L}}\beta_T)\alpha_T^2}{\rho_{\mathcal{L}}\beta_T}\E\left\|\nabla f(x_k)\right\|^2+\frac{n\alpha_T^2\sigma_{\ell}^2}{\gamma_T}\cr
	=&(1\!+\!\rho_{\mathcal{L}}\beta_T)(1\!-\!\rho_{\mathcal{L}}\beta_T)^2\E\|Y_k\|^2+nr\beta_T^2(\Delta^2\!+\!\sigma_k^2)\cr
	&\!\!+\!\!\frac{(1\!\!+\!\!\rho_{\mathcal{L}}\beta_T)\alpha_T^2}{\rho_{\mathcal{L}}\beta_T}\E\|\nabla \!f\!(x_k)\!\!-\!\!\nabla \!f\!(\bar{x}_k)\!\!+\!\!\nabla \!f\!(\bar{x}_k)\!\|^2\!\!+\!\!\frac{n\alpha_T^2\sigma_{\ell}^2}{\gamma_T}\!.~~~~~
\end{align}

Note that for any $m\ge1$ and $\mathbf{a}_1,\mathbf{a}_2,\dots,\mathbf{a}_m\in\mathbb{R}^{r}$, the following inequality holds:\vspace{-0.5em}
\begin{align}\label{eq: abs-triangle inequality}
	\|\mathbf{a}_1\!+\!\mathbf{a}_2\!+\!\dots\!+\!\mathbf{a}_m\!\|^2\!\le\! m(\|\mathbf{a}_1\!\|^2\!+\!\|\mathbf{a}_2\!\|^2\!+\!\dots\!+\!\|\mathbf{a}_m\!\|^2).
\end{align}
{\vskip -3pt}\noindent Then, by letting $m=2$ in \eqref{eq: abs-triangle inequality}, $\|\nabla \!f(x_k)\!\!-\!\!\nabla \!f(\bar{x}_k)\!\!+\!\!\nabla \!f(\bar{x}_k)\|^2$ in \eqref{eq:consensus ,1.5} can be rewritten as\vspace{-0.5em}
\begin{align}\label{eq:consensus ,2}
	&\left\|\nabla f(x_k)\right\|^2\cr
	\leq &2\left\|\nabla f(x_k)-\nabla f(\bar{x}_k)\right\|^2+2\left\|\nabla f(\bar{x}_k)\right\|^2\cr
	\noalign{\vskip -3pt}
	=&2\!\sum_{i=1}^{n}\!\left\|\!\nabla f_i(x_{i,k})\!-\!\nabla f_i(\bar{x}_k)\!\right\|^2\!+\!2\!\sum_{i=1}^{n}\left\|\!\nabla f_i(\bar{x}_k)\!\right\|^2\!.
\end{align}

{\vskip -6pt}\indent By Assumption \ref{assumption 2: cost functions}(i), for any $x,y\in\mathbb{R}^r$, we have\vspace{-0.5em}
\begin{align*}
	&\|\nabla f_i(x)-\nabla f_i(y)\|=\|\E \nabla\ell(x,\xi_i)-\E\nabla\ell(y,\xi_i)\|\cr
	\leq& \E\| \nabla\ell(x,\xi_i)-\nabla\ell(y,\xi_i)\|\leq L_1\|x-y\|.
\end{align*}{\vskip -5pt}\noindent Then, it can be seen that $\|\nabla f_i(x_{i,k})-\nabla f_i(\bar{x}_k)\|\leq L_1\|x_{i,k}-\bar{x}_k\|$. Since $\|Y_k\|^2=\|(W\otimes I_{r})x_k\|^2=\sum_{i=1}^{n}\|x_{i,k}-\bar{x}_k\|^2$, $\sum_{i=1}^{n}\left\|\nabla f_i(x_{i,k})-\nabla f_i(\bar{x}_k)\right\|^2$ in  \eqref{eq:consensus ,2} can be rewritten as\vspace{-0.5em}
\begin{align}\label{eq: gradient error,1}
	\hspace{-1.5em}\sum_{i=1}^{n}\!\left\|\nabla f_i(x_{i,k})\!\!-\!\!\nabla f_i(\bar{x}_k)\!\right\|^2\!\!\leq\! L_1^2\!\sum_{i=1}^{n}\!\left\|\! x_{i,k}\!\!-\!\!\bar{x}_k\!\right\|^2\!\!=\!\!L_1^2\|Y_k\|^2\!.
\end{align}
{\vskip -6pt}\noindent By Assumption \ref{assumption 2: cost functions}(ii) and Lemma \ref{lemma a1}(ii), $\|\nabla f_i(\bar{x}_k)\|^2\leq 2L_1(f_i(\bar{x}_k)-f_i^*)$, we have\vspace{-0.5em}
\begin{align}\label{eq: gradient error,2}
	\sum_{i=1}^{n}\left\|\nabla f_i(\bar{x}_k)\right\|^2\leq 2L_1\sum_{i=1}^{n}(f_i(\bar{x}_k)-f_i^*).
\end{align}{\vskip -3pt}\noindent Thus, substituting \eqref{eq: gradient error,1} and \eqref{eq: gradient error,2} into \eqref{eq:consensus ,2} gives
\begin{align}\label{eq: gradient error,3}
	&\left\|\nabla f(x_k)-\nabla f(\bar{x}_k)+\nabla f(\bar{x}_k)\right\|^2\cr
	\le&2L_1^2\|Y_k\|^2+4L_1\left(\sum_{i=1}^{n}f_{i}(\bar{x}_k)-f_i^*\right).
\end{align}\vspace{-8pt}

Note that by Assumption \ref{assumption 2: cost functions}(ii), each cost function $f_i(x)$ has the minimum $f_i^*$. Then, the global cost function $F(x)$ has the global minimum $F^*=\min_{x\in\mathbb{R}^{r}}F(x)$. Let $M^*=F^*-\frac{1}{n}\sum_{i=1}^{n}f_i^*$. Then, \eqref{eq: gradient error,3} can be rewritten as $\|\nabla f(x_k)-\nabla f(\bar{x}_k)+\nabla f(\bar{x}_k)\|^2\leq2L_1^2\|Y_k\|^2+4L_1(\sum_{i=1}^{n}f_{i}(\bar{x}_k)-f_i^*)=2L_1^2\|Y_k\|^2+4nL_1(F(\bar{x}_k)-F^*)+4nL_1M^*$. This together with \eqref{eq:consensus ,1.5} implies
\begin{align}\label{eq:consensus ,4}
	&\E\|Y_{k+1}\|^2\leq\left(\!1\!-\!\rho_{\mathcal{L}}\beta_T+\frac{2(1\!+\!\rho_{\mathcal{L}}\beta_T)\alpha_T^2L_1^2}{\rho_{\mathcal{L}}\beta_T}\!\right)\E\|Y_k\|^2\nonumber\\
	&+\frac{4n(1+\rho_{\mathcal{L}}\beta_T)\alpha_T^2L_1}{\rho_{\mathcal{L}}\beta_T}\E(F(\bar{x}_k)-F^*)+\frac{n\alpha_T^2\sigma_{\ell}^2}{\gamma_T}\cr
	&+nr\beta_T^2(\Delta^2+\sigma_k^2)+\frac{4n(1+\rho_{\mathcal{L}}\beta_T)\alpha_T^2L_1M^*}{\rho_{\mathcal{L}}\beta_T}.
\end{align}

\textbf{Step 2:} We next focus on the term $F(\bar{x}_k)-F^*$. Multiplying both sides of \eqref{eq:compact form} by $\frac{1}{n}(\mathbf{1}_{n}^\top\otimes I_{r})$ implies
\begin{align}\label{eq: optimization, 1}
	\hspace{-0.5em}\smash{\bar{x}_{k+1}\!=\!\bar{x}_k\!-\!\alpha_T\overline{\nabla f(x_k)}\!-\!\alpha_T\bar{w}_k\!+\!\frac{\beta_T}{n}(\mathbf{1}_n^\top\!\!\otimes\! I_{r}\!)(\!e_k\!+\!d_k\!).}
\end{align}
Then by \eqref{eq: optimization, 1} and Lemma \ref{lemma a1}(i), we can derive that\vspace{-0.5em}
\begin{align}\label{eq: optimization, 2}
	&F(\bar{x}_{k+1})-F^*\cr
	\noalign{\vskip -3pt}
	\leq& \left(F(\bar{x}_k)-F^*\right)+\frac{L_1}{2}\|\bar{x}_{k+1}-\bar{x}_{k}\|^2+\langle \nabla F(\bar{x}_k),\bar{x}_{k+1}-\bar{x}_{k} \rangle\cr
	=&\left(F(\bar{x}_k)-F^*\right)+\frac{L_1}{2}\|\alpha_T\overline{\nabla f(x_k)}-\frac{\beta_T}{n}\left(\mathbf{1}_n^\top\otimes I_{r}\right)\left(e_k+d_k\right)\cr
	\noalign{\vskip -3pt}
	&+\alpha_T\bar{w}_k\|^2-\langle \nabla F(\bar{x}_k), -\frac{\beta_T}{n}\left(\mathbf{1}_n^\top\otimes I_{r}\right)\left(e_k+d_k\right)\cr
	&+\alpha_T\overline{\nabla f(x_k)}+\alpha_T\bar{w}_k\rangle.
\end{align}
{\vskip -3pt}\noindent By \eqref{eq:expectation of nk}, \eqref{eq:expectation of wk} and \eqref{eq:expectation of ek}, taking mathematical expectation of \eqref{eq: optimization, 2} gives\vspace{-0.5em}
\begin{align}\label{eq: optimization, 3}
	&\E\left(F(\bar{x}_{k+1})-F^*\right)\cr
	\noalign{\vskip -3pt}
	\leq&\E(F(\bar{x}_k)-F^*)-\alpha_T\E\left\langle\nabla F(\bar{x}_k),\overline{\nabla f(x_k)}\right\rangle\cr
	\noalign{\vskip -7pt}
	&+\frac{L_1}{2}\E\|\alpha_T\overline{\nabla f(x_k)}\!-\!\frac{\beta_T}{n}\left(\mathbf{1}_n^\top\!\otimes\! I_{r}\right)\left(e_k\!+\!d_k\right)\!+\!\alpha_T\bar{w}_k\|^2\cr
	=&\smash{\E(F(\bar{x}_k)-F^*)-\alpha_T\E\left\langle\nabla F(\bar{x}_k),\overline{\nabla f(x_k)}\right\rangle}\cr
	\noalign{\vskip -4pt}
	&+\frac{\beta_T^2L_1}{2n^2}\E\left(\|\left(\mathbf{1}_n^\top\otimes I_{r}\right)e_k\|^2+\|\left(\mathbf{1}_n^\top\otimes I_{r}\right)d_k\|^2\right)\cr
	&+\frac{\alpha_T^2L_1}{2}\E\left\|\overline{\nabla f(x_k)}\right\|^2+\frac{\alpha_T^2L_1}{2}\E\|\bar{w}_k\|^2.
\end{align}

{\vskip -3pt}\indent Note that $\|(\mathbf{1}_n^\top\otimes I_{r})d_k\|^2=\|\sum_{i=1}^{n} d_{i,k}\|^2\leq n\|d_k\|^2$, $\|(\mathbf{1}_n^\top\otimes I_{r})e_k\|^2=\|\sum_{i=1}^{n} e_{i,k}\|^2\leq n\|e_k\|^2$, $\|\bar{w}_k\|^2=\|\frac{1}{n}\sum_{i=1}^{n} w_{i,k}\|^2\le\frac{1}{n}\sum_{i=1}^{n}\|w_{i,k}\|^2$. Then, by \eqref{eq:variance of nk}, \eqref{eq:variance of wk} and \eqref{eq:variance of ek}, \eqref{eq: optimization, 3} can be rewritten as
\begin{align}\label{eq: optimization, 3.5}
	\hspace{-1em}&\E\left(F(\bar{x}_{k+1})-F^*\right)\cr
	\hspace{-1em}\leq&\E(F(\bar{x}_k)-F^*)-\alpha_T\E\left\langle\nabla F(\bar{x}_k),\overline{\nabla f(x_k)}\right\rangle\cr
	&+\!\frac{\alpha_T^2L_1}{2}\E\left\|\overline{\nabla f(x_k)}\right\|^2\!\!+\!\frac{\beta_T^2nrL_1}{2}(\Delta^2\!+\!\sigma_k^2)\!+\!\frac{\alpha_T^2\sigma_{\ell}^2L_1}{2\gamma_T}.
\end{align}

{\vskip -3pt}\indent Note that $\langle \mathbf{a},\mathbf{b}\rangle=\frac{1}{2}\|\mathbf{a}\|^2+\frac{1}{2}\|\mathbf{b}\|^2-\frac{1}{2}\|\mathbf{a}-\mathbf{b}\|^2$ for any $\mathbf{a}$, $\mathbf{b}\in\mathbb{R}^{r}$. Then, $-\alpha_T\langle\nabla F(\bar{x}_k),\overline{\nabla f(x_k)}\rangle$ in \eqref{eq: optimization, 3.5} can be \mbox{rewritten as}\vspace{-0.5em}
\begin{align}\label{eq: error of consensus,1}
	&-\alpha_T\left\langle\nabla F(\bar{x}_k),\overline{\nabla f(x_k)}\right\rangle\cr
	\noalign{\vskip -2pt}
	=&-\!\!\frac{\alpha_T}{2}\!\left\|\nabla F(\bar{x}_k)\right\|^2\!\!-\!\!\frac{\alpha_T}{2}\!\left\|\overline{\nabla f(x_k)}\right\|^2\!\!+\!\!\frac{\alpha_T}{2}\!\left\|\nabla F(\bar{x}_k)\!\!-\!\!\overline{\nabla f(x_k)}\right\|^2\cr
	\le&-\frac{\alpha_T}{2}\left\|\nabla F(\bar{x}_k)\right\|^2+\frac{\alpha_T}{2}\left\|\nabla F(\bar{x}_k)-\overline{\nabla f(x_k)}\right\|^2.
\end{align}
{\vskip -4pt}\indent By letting $m=n$ in \eqref{eq: abs-triangle inequality}, $\|\nabla F(\bar{x}_k)-\overline{\nabla f(x_k)}\|^2$ in \eqref{eq: error of consensus,1} can be rewritten as\vspace{-0.5em}
\begin{align}\label{eq: error of consensus,1.5}
	\hspace{-1em}\left\|\nabla F(\bar{x}_k)-\overline{\nabla f(x_k)}\right\|^2=&\left\|\!\frac{1}{n}\!\sum_{i=1}^{n}\!(\nabla f_i(\bar{x}_k)\!-\!\nabla f_i(x_{i,k})\!)\!\right\|^2\nonumber\\
	\leq&\frac{1}{n}\!\sum_{i=1}^{n}\!\left\|\nabla f_i(\bar{x}_k)\!-\!\nabla f_i(x_{i,k})\right\|^2.
\end{align}
{\vskip -4pt}\noindent Thus, by \eqref{eq: gradient error,1}, \eqref{eq: error of consensus,1.5} can be rewritten as\vspace{-0.5em}
\begin{align}\label{eq: error of consensus,2}
	\left\|\nabla F(\bar{x}_k)\!-\!\overline{\nabla f(x_k)}\right\|^2\le\frac{L_1^2}{n}\|Y_k\|^2.
\end{align}
{\vskip -4pt}\noindent Substituting \eqref{eq: error of consensus,1} and \eqref{eq: error of consensus,2} into \eqref{eq: optimization, 3.5} implies\vspace{-0.5em}
\begin{align*}
	&\E(F(\bar{x}_{k+1})-F^*)\cr
	\noalign{\vskip -3pt}
	\leq&\E(F(\bar{x}_k)-F^*)-\frac{\alpha_T}{2}\E\|\nabla F(\bar{x}_k)\|^2+\frac{\alpha_TL_1^2}{2n}\E\|Y_k\|^2\cr
	\noalign{\vskip -3pt}
	&+\frac{\alpha_T^2L_1}{2}\E\left\|\overline{\nabla f(x_k)}-\nabla F(\bar{x}_k)+ \nabla F(\bar{x}_k)\right\|^2
\end{align*}
	\begin{align}\label{eq: optimization, 4}
	&+\frac{\beta_T^2nrL_1}{2}\left(\Delta^2+\sigma_k^2\right)+\frac{\alpha_T^2\sigma_{\ell}^2L_1}{2\gamma_T}.
\end{align}
{\vskip -3pt}\indent Furthermore, by letting $m=2$ in \eqref{eq: abs-triangle inequality} and using \eqref{eq: error of consensus,2}, $\| \overline{\nabla f(x_k)}-\nabla F(\bar{x}_k)+ \nabla F(\bar{x}_k)\|^2$ in \eqref{eq: optimization, 4} can be \mbox{rewritten as}\vspace{-0.2em}
\begin{align}\label{eq: optimization, 5}
	&\left\| \overline{\nabla f(x_k)}-\nabla F(\bar{x}_k)+ \nabla F(\bar{x}_k)\right\|^2\cr
	\le&2\left\| \overline{\nabla f(x_k)}-\nabla F(\bar{x}_k)\right\|^2+2\left\| \nabla F(\bar{x}_k)\right\|^2\cr
	\le&\frac{2L_1^2}{n}\|Y_k\|^2+2\left\| \nabla F(\bar{x}_k)\right\|^2.
\end{align}
{\vskip -6pt}\noindent By letting $m=n$ in \eqref{eq: abs-triangle inequality} and using \eqref{eq: gradient error,2}, $\| \nabla F(\bar{x}_k)\|^2$ in \eqref{eq: optimization, 5} can be rewritten as\vspace{-0.5em}
\begin{align}\label{eq: error of consensus,3}
	\left\|\nabla F(\bar{x}_k)\right\|^2\leq&\frac{1}{n}\sum_{i=1}^{n}\left\|\nabla f_i(\bar{x}_k)\right\|^2\cr
	\leq&\!\frac{2L_1}{n}\!\sum_{i=1}^{n}\!\left(f_i(\bar{x}_k)\!-\!f_i^*\right)\cr
	=&2L_1\left(F(\bar{x}_k)\!-\!F^*\right)\!+\!2L_1M^*\!.\!
\end{align}
{\vskip -6pt}\noindent Thus, substituting \eqref{eq: optimization, 5}-\eqref{eq: error of consensus,3} into \eqref{eq: optimization, 4} implies\vspace{-0.5em}
\begin{align}\label{eq: optimization, 6}
	&\E\left(F(\bar{x}_{k+1})-F^*\right)\cr
	\leq&\left(1+2\alpha_T^2L_1^2 \right)\E(F(\bar{x}_k)-F^*)\cr
	&-\frac{\alpha_T}{2}\E\|\nabla F(\bar{x}_k)\|^2+\frac{\alpha_TL_1^2(1+2\alpha_TL_1)}{2n}\E\|Y_k\|^2\cr
	&+\frac{\alpha_T^2\sigma_{\ell}^2L_1}{2\gamma_T}+\frac{\beta_T^2 nrL_1}{2}(\Delta^2+\sigma_k^2)+2\alpha_T^2L_1^2M^*\cr
	\leq&\left(1+2\alpha_T^2L_1^2 \right)\E(F(\bar{x}_k)-F^*)\cr
	&+\frac{\alpha_TL_1^2(1+2\alpha_TL_1)}{2n}\E\|Y_k\|^2+\frac{\alpha_T^2\sigma_{\ell}^2L_1}{2\gamma_T}\cr
	&+\frac{\beta_T^2 nrL_1}{2}(\Delta^2+\sigma_k^2)+2\alpha_T^2L_1^2M^*.
\end{align}

Let\vspace{-0.6em}
\begin{align}
	\theta_1=&\max\{1\!+\!2\alpha_T^2L_1^2\!+\!\frac{4n(1+\rho_{\mathcal{L}}\beta_T)\alpha_T^2L}{\rho_{\mathcal{L}}\beta_T},\cr
	&1\!-\!\rho_{\mathcal{L}}\beta_T\!\!+\!\!\frac{\alpha_TL_1^2\!(1\!+\!2\alpha_TL_1)}{2n}\!+\!\frac{2(1\!\!+\!\!\rho_{\mathcal{L}}\beta_T)\alpha_T^2\!L_1^2}{\rho_{\mathcal{L}}\beta_T}\!\},~~~~\label{eq: theta 1}\\ \theta_{k,2}=&\frac{(L_1+2)nr\beta_T^2}{2}(\Delta^2+\sigma_k^2)+\frac{\alpha_T^2\sigma_{\ell}^2(2n+L_1)}{2\gamma_T}\nonumber\\
	&+2\alpha_T^2L_1^2M^*+\frac{4n(1+\rho_{\mathcal{L}}\beta_T)\alpha_T^2L_1M^*}{\rho_{\mathcal{L}}\beta_T}.\label{eq: theta 2}
\end{align}{\vskip -5pt}\noindent Then, summing \eqref{eq:consensus ,4} and \eqref{eq: optimization, 6} implies\vspace{-0.5em}
\begin{align}\label{eq: convergence rate,1}
	&\E(\|Y_{k+1}\|^2+F(\bar{x}_{k+1})-F^*)\cr
	\le&\theta_1 \E(\|Y_{k}\|^2+F(\bar{x}_{k})-F^*)+\theta_{k,2}.
\end{align}
{\vskip -5pt}\noindent By iteratively computing \eqref{eq: convergence rate,1}, the following inequality holds:\vspace{-0.5em}
\begin{align}\label{eq: convergence rate,2}
		&\E(\|Y_{T+1}\|^2+F(\bar{x}_{T+1})-F^*)\cr
		\noalign{\vskip -5pt}
		\le&\theta_1^{T+1}(\|Y_0\|^2\!+\!F(\bar{x}_0)\!-\!F^*)+\sum_{k=0}^{T}\theta_1^{T-k}\theta_{k,2}.
\end{align}{\vskip -5pt}

\textbf{Step 3:} At this step, we prove that there exists $G_1\geq0$ such that $\E(F(\bar{x}_T)-F^*)\leq G_1$ for any $T=0,1,\dots$. Note that $2\alpha_T^2L_1^2=O(\frac{1}{(T+1)^{2u}})$ and $\frac{4n(1+\rho_{\mathcal{L}}\beta_T)\alpha_T^2L_1}{\rho_{\mathcal{L}}\beta_T}=O(\frac{1}{(T+1)^{2u-v}})$ holds for any $T=0,1,\dots$. Then, by $2u-v>1$ in Assumption \ref{assumption 4: step sizes}, it can be seen that for any $T=0,1,\dots$,\vspace{-0.5em}
\begin{align*}
	&\left(\!1\!+\!2\alpha_T^2\!L_1^2\!+\!\frac{4n(1\!+\!\rho_{\mathcal{L}}\beta_T)\alpha_T^2\!L_1}{\rho_{\mathcal{L}}\beta_T}\!\right)^{T\!+\!1}\cr
	=&\left(\!1\!+\!O\left(\!\frac{1}{(T+1)^{2u-v}}\!\right)\!\right)^{T\!+\!1}
\end{align*}
	\begin{align}\label{eq59}
	=&\exp\left((T+1)\ln\left(1+O\left(\frac{1}{(T+1)^{2u-v}}\right)\right)\right)\cr
	=&\exp\left(O\left(\frac{1}{(T+1)^{2u-v-1}}\right)\right).
\end{align}{\vskip -7pt}
Note that $\ln$$ (1+x)$$\leq$$ x$ for any $x$$>$$-1$, 
and by $\frac{1}{2}$$+$$\max\{w,0\}$$<$$v$$<$$u$ in Assumption \ref{assumption 4: step sizes}, there exists a positive integer $T_0$ such that $1-\rho_{\mathcal{L}}\beta_T+\frac{\alpha_TL_1^2\!(1\!+\!2\alpha_TL_1)}{2n}+\frac{2(1\!+\!\rho_{\mathcal{L}}\beta_T)\alpha_T^2\!L_1^2}{\rho_{\mathcal{L}}\beta_T}\leq1-\frac{\rho_{\mathcal{L}}\beta_T}{2}$ for any $T=T_0, T_0+1,\dots$. Then, it can be seen that for any $T=T_0, T_0+1,\dots$,\vspace{-0.3em}
\begin{align}\label{eq60}
	&\left(1\!-\!\rho_{\mathcal{L}}\beta_T\!+\!\frac{\alpha_TL_1^2\!(1\!+\!2\alpha_TL_1)}{2n}\!+\!\frac{2(1\!+\!\rho_{\mathcal{L}}\beta_T)\alpha_T^2\!L_1^2}{\rho_{\mathcal{L}}\beta_T}\right)^{T+1}\nonumber\\
	\leq&\left(1\!-\!\frac{\rho_{\mathcal{L}}\beta_T}{2}\right)^{T\!+\!1}=\exp\left(\!(T\!+\!1)\ln\!\left(\!1\!-\!\frac{\rho_{\mathcal{L}}a_2}{2(T+1)^{v}}\right)\!\right)\nonumber\\
	\leq&\exp\left(-\frac{\rho_{\mathcal{L}}a_2}{2}(T+1)^{1-v}\right)\cr
	\leq&\exp\left(-\frac{\rho_{\mathcal{L}}a_2}{2}T_0^{1-v}\right).
\end{align}{\vskip -3pt}\noindent Thus, for any $T=0,1,\dots$, we have\vspace{-0.5em}
\begin{align}\label{eq61}
	&\left(1\!-\!\rho_{\mathcal{L}}\beta_T\!\!+\!\!\frac{\alpha_TL_1^2\!(1\!+\!2\alpha_TL_1)}{2n}\!+\!\frac{2(1\!+\!\rho_{\mathcal{L}}\beta_T)\alpha_T^2\!L^2}{\rho_{\mathcal{L}}\beta_T}\right)^{T+1}\nonumber\\
	\leq&\max\{\exp\left(-\frac{\rho_{\mathcal{L}}a_2}{2}T_0^{1-v}\right),\cr
	&1\!\!-\!\!\rho_{\mathcal{L}}a_2\!\!+\!\!\frac{a_1\!L_1^2(2a_1\!L_1\!\!+\!\!1)}{2n}\!\!+\!\!\frac{2(1\!\!+\!\!\rho_{\mathcal{L}}a_2)a_1^2\!L_1^2}{\rho_{\mathcal{L}}a_2},\dots,\cr
	&1\!\!-\!\!\frac{\rho_{\mathcal{L}}a_2}{T_0^{v}}\!\!+\!\!\frac{a_1\!L_1^2(2a_1\!L_1\!\!+\!\!T_0^{u})}{2nT_0^{2u}}\!\!+\!\!\frac{2(T_0^{v}\!\!+\!\!\rho_{\mathcal{L}}a_2)a_1^2\!L^2}{\rho_{\mathcal{L}}a_2T_0^{2u}}\!\}.
\end{align}Hence, \eqref{eq59} together with \eqref{eq61} implies that there exists $G_0>1$ such that for any $T=0,1,\dots$,
\begin{align}\label{eq61.5}
1<\theta_{1}^{T+1}\leq G_0.
\end{align}{\vskip -4pt}

When $w\le0$, $\sigma_{k}$ is decreasing, and then $\sigma_{k}\le\sigma_{0}$ for any $k=0,\dots, T$. When $w>0$, $\sigma_{k}$ is increasing, and then $\sigma_{k}\le\sigma_{T}$ for any $k=0,\dots, T$. As a result, $\sigma_{k}\leq \max\{\sigma_{0},\sigma_{T}\}$ for any $k=0,\dots, T$. Hence, by the definition of $\theta_{k,2}$ in \eqref{eq: theta 2}, $\theta_{k,2}\le\max\{\theta_{0,2},\theta_{T,2}\}$ for any $k=0,\dots, T$. This helps us to obtain that\vspace{-0.5em}
\begin{align}\label{eq62}
	\sum_{k=0}^{T}\theta_1^{T-k}\theta_{k,2}\leq&\sum_{k=0}^{T}\theta_1^{T+1}\theta_{k,2}\cr
	\noalign{\vskip -2pt}
	\le&(T+1)\max\{\theta_{0,2},\theta_{T,2}\}\theta_1^{T+1}.
\end{align}{\vskip -5pt}\noindent Note that\vspace{-0.4em}
\begin{align}\label{eq63}
	&\max\{\theta_{0,2},\theta_{T,2}\}\cr
	=&\frac{(L_1+2)nr\beta_T^2}{2}(\Delta^2+\max\{\sigma_0^2,\sigma_T^2\})\nonumber\\
	&+\frac{\alpha_T^2\sigma_{\ell}^2(2n+L_1)}{2\gamma_T}+2\alpha_T^2L_1^2M^*\cr
	&+\frac{4n(1+\rho_{\mathcal{L}}\beta_T)\alpha_T^2L_1M^*}{\rho_{\mathcal{L}}\beta_T}\cr
	=&O\left(\frac{1}{(T+1)^{2v-2\max\{w,0\}}}+\frac{1}{(T+1)^{2u-v}}\right).
\end{align}{\vskip -4pt}\noindent Then, by $2u-v>1$ and $\frac{1}{2}+\max\{w,0\}<v$ in Assumption~\ref{assumption 4: step sizes}, substituting \eqref{eq63} into \eqref{eq62} implies\vspace{-0.5em}
\begin{align}\label{eq64}
	\sum_{k=0}^{T}\!\!\theta_1^{T\!-\!k}\theta_{k,2}\!\!=\!\!O\!\!\left(\!\frac{1}{(T\!\!+\!\!1)^{2v\!-\!2\max\{w\!,0\}\!-\!1}}\!\!+\!\!\frac{1}{(T\!\!+\!\!1)^{2u\!-\!v\!-\!1}}\!\right).
\end{align}
{\vskip -4pt}\noindent Thus, by \eqref{eq64} there exists $G_0^\prime$$>$$0$ such that for any $T$$=$$0$$,$$1$$,$$\dots$,\vspace{-0.6em}
\begin{align}\label{eq64.1}
	\sum_{k=0}^{T}\!\!\theta_{1}^{T\!-\!k}\theta_{k,2}\leq G_0^\prime.
\end{align}
{\vskip -5pt}\noindent By \eqref{eq: convergence rate,2}, \eqref{eq61.5} and \eqref{eq64.1} we have\vspace{-0.5em}
\begin{align*}
	&\E(\|Y_{T+1}\|^2+F(\bar{x}_{T+1})-F^*)\cr
	\noalign{\vskip -5pt}
	\leq&\theta_{1}^{T+1}(\|Y_0\|^2\!+\!F(\bar{x}_0)\!-\!F^*)+\sum_{k=0}^{T}\theta_{1}^{T-k}\theta_{k,2}\cr
	\noalign{\vskip -3pt}
	\leq& G_0(\|Y_0\|^2\!+\!F(\bar{x}_0)\!-\!F^*)+G_0^\prime.
\end{align*}
{\vskip -3pt}\noindent Let $G_1=G_0(\|Y_0\|^2\!+\!F(\bar{x}_0)\!-\!F^*)+G_0^\prime$. Then, there exists $G_1\geq0$ such that $\E(F(\bar{x}_T)-F^*)\leq G_1$ for any $T=0,1,\dots$.

{\bf Step 4:} At this step, we prove $\lim_{T\to\infty}\E\|Y_{T+1}\|^2=0$. By Step 3, since there exists $G_1\geq0$ such that $\E(F(\bar{x}_T)-F^*)\leq G_1$ for any $T=0,1,\dots$, by \eqref{eq:consensus ,4} we have\vspace{-0.6em}
\begin{align}\label{eq66}
	\E\|Y_{k+1}\|^2\leq&\left(\!1\!-\!\rho_{\mathcal{L}}\beta_T+\frac{2(1\!+\!\rho_{\mathcal{L}}\beta_T)\alpha_T^2L_1^2}{\rho_{\mathcal{L}}\beta_T}\!\right)\E\|Y_k\|^2\nonumber\\
	&+\frac{4n(1+\rho_{\mathcal{L}}\beta_T)\alpha_T^2L_1(G_1+M^*)}{\rho_{\mathcal{L}}\beta_T}+\frac{n\alpha_T^2\sigma_{\ell}^2}{\gamma_T}\nonumber\\
	&+nr\beta_T^2(\Delta^2\!+\!\sigma_k^2).
\end{align}
Let \vspace{-0.6em}
\begin{align}
	\theta_3=&1\!-\!\rho_{\mathcal{L}}\beta_T+\frac{2(1\!+\!\rho_{\mathcal{L}}\beta_T)\alpha_T^2L_1^2}{\rho_{\mathcal{L}}\beta_T},\label{theta3}\\
	\theta_{k,4}=&\frac{4n(1+\rho_{\mathcal{L}}\beta_T)\alpha_T^2L_1(G_1+M^*)}{\rho_{\mathcal{L}}\beta_T}+\frac{n\alpha_T^2\sigma_{\ell}^2}{\gamma_T}\cr
	&+nr\beta_T^2(\Delta^2\!+\!\sigma_k^2).\label{theta4}
	\end{align}
{\vskip -3pt}\noindent Then, substituting \eqref{theta3} and \eqref{theta4} into \eqref{eq66} and iteratively computing \eqref{eq66} gives\vspace{-0.7em}
\begin{align}\label{eq69}
	\E\|Y_{k+1}\|^2\leq\theta_3^{k+1}\|Y_0\|^2+\sum_{m=0}^{k}\theta_3^{k-m}\theta_{m,4}.
\end{align}

{\vskip -4pt}Note that by the definition of $\theta_3$ in \eqref{theta3} and $2u-v>1$, $\frac{1}{2}+\max\{w,0\}<v<u<1$ in Assumption \ref{assumption 4: step sizes}, we have\vspace{-0.4em}
\begin{align}\label{eq70}
	\frac{1}{1-\theta_3}=\frac{1}{\rho_{\mathcal{L}}\beta_T\!-\!\frac{2(1\!+\!\rho_{\mathcal{L}}\beta_T)\alpha_T^2L_1^2}{\rho_{\mathcal{L}}\beta_T}}=O\!\left((T+1)^{v}\right),
\end{align}\vspace{-1em}
\begin{align}\label{eq70.1}
	\hspace{-2em}\max\{\theta_{0,4},\theta_{T,4}\}
	=&\frac{4n(1+\rho_{\mathcal{L}}\beta_T)\alpha_T^2L_1(G_1+M^*)}{\rho_{\mathcal{L}}\beta_T}\cr
	&+\frac{n\alpha_T^2\sigma_{\ell}^2}{\gamma_T}+nr\beta_T^2(\Delta^2\!+\!\max\{\sigma_T^2,\sigma_0^2\})\cr
	\noalign{\vskip -3pt}
	=&O\left(\!\!\frac{1}{(T\!+\!1)^{2u\!-\!v}}\!+\!\frac{1}{(T\!+\!1)^{2v\!-\!2\max\{w,0\}}}\!\!\right).
\end{align}
{\vskip -3pt}\noindent Moreover, by the definition of $\theta_{k,4}$ in \eqref{theta4}, $\theta_{k,4}\leq\max\{\theta_{0,4},$ $\theta_{T,4}\}$ for any $k=0,\dots,T$. Then, it follows from \eqref{eq70} and \eqref{eq70.1} that\vspace{-0.5em}
\begin{align}\label{eq70.2}
	&\sum_{k=0}^{T}\theta_3^{T-k}\theta_{k,4}\leq\max\{\theta_{0,4},\theta_{T,4}\}\sum_{k=0}^{T}\theta_3^{T-k}\cr
	=&\max\{\theta_{0,4},\theta_{T,4}\}\frac{1-\theta_3^{T+1}}{1-\theta_3}
	=O\left(\frac{\max\{\theta_{0,4},\theta_{T,4}\}}{1-\theta_3}\right)\cr
	=&O\left(\frac{1}{(T\!+\!1)^{2u-2v}}+\frac{1}{(T\!+\!1)^{v-2\max\{w,0\}}}\right)\!.
\end{align}
{\vskip -2pt}\noindent  Meanwhile, by \eqref{eq60} we have\vspace{-0.3em}
\begin{align*}
	\theta_3^{T+1}\leq&\left(1\!-\!\rho_{\mathcal{L}}\beta_T\!\!+\!\!\frac{\alpha_TL_1^2\!(1\!\!+\!\!2\alpha_TL_1)}{2n}\!\!+\!\!\frac{2(1\!\!+\!\!\rho_{\mathcal{L}}\beta_T)\alpha_T^2\!L_1^2}{\rho_{\mathcal{L}}\beta_T}\right)^{\!\!T\!+\!1}\nonumber\\
	=&O((1-\frac{\rho_{\mathcal{L}}\beta_T}{2})^{T+1})
\end{align*}
	\begin{align}\label{eq70.3}
	=&O\left(\exp\left((T+1)\ln\left(1-\frac{\rho_{\mathcal{L}}\beta_T}{2}\right)\right)\right)\cr
	\noalign{\vskip -2pt}
	=&O\left(\exp\left(-\frac{\rho_{\mathcal{L}}a_2}{2}(T\!+\!1)^{1-v}\right)\right).
\end{align}

{\vskip -2pt}Let $k=T$ in \eqref{eq69}. Then, substituting   \eqref{eq70.2} and \eqref{eq70.3} into \eqref{eq69} implies $\E\|Y_{T+1}\|^2$$=$$ O(\exp(-\frac{\rho_{\mathcal{L}}a_2}{2}(T\!+\!1)^{1-v}))$$+$$O(\frac{1}{(T\!+\!1)^{2u-2v}}$ $+\frac{1}{(T\!+\!1)^{v-2\max\{w,0\}}})=O(\frac{1}{(T\!+\!1)^{2u-2v}}\!+\!\frac{1}{(T\!+\!1)^{v\!-\!2\max\!\{w,0\}}})$. Hence, we have $\lim_{T\to\infty}\E\|Y_{T+1}\|^2=0$.

{\bf Step 5:} At this step, we give the estimation of $\sum_{k=0}^{T}\E\|Y_k\|^2$ for any $T=0,1,\dots$. By defining $\sum_{k=1}^{0}\!\sum_{m=0}^{k}\!\theta_3^{k-m}\theta_{m,4}=0$, summing \eqref{eq69} from $k=0$ to $T$ gives\vspace{-0.6em}
\begin{align}\label{eq70.5}
  \sum_{k=0}^{T}\!\E\|Y_{k}\|^2\!\leq\!\sum_{k=0}^{T}\theta_3^{k}\|Y_0\|^2\!+\!\sum_{k=1}^{T}\!\sum_{m=0}^{k}\!\theta_3^{k-m}\theta_{m,4}.
\end{align}{\vskip -5pt}\noindent Then, it follows from \eqref{eq70} that\vspace{-0.6em}
\begin{align}\label{eq71}
 \sum_{k=0}^{T}\theta_3^{k}\|Y_0\|^2=\frac{1-\theta_3^{T+1}}{1-\theta_3}\|Y_0\|^2=O\left((T\!+\!1)^{v}\right).
\end{align}
{\vskip -5pt}\noindent Moreover, by \eqref{eq70}-\eqref{eq70.2}, we have\vspace{-0.4em}
\begin{align}\label{eq72}
  &\sum_{k=1}^{T}\!\sum_{m=0}^{k}\!\theta_3^{k\!-\!m}\!\theta_{m,4}\leq\max\{\theta_{0,4},\theta_{T,4}\}\sum_{k=1}^{T}\sum_{m=0}^{k}\theta_3^{k\!-\!m}\cr
  \noalign{\vskip -5pt}
  =&\max\{\theta_{0,4},\theta_{T,4}\}\sum_{k=1}^{T}\frac{1-\theta_3^{k+1}}{1-\theta_3}\cr
  =&O\!\left(\!\frac{T\max\{\theta_{0,4},\theta_{T,4}\}}{1-\theta_3}\!\right)\cr
  =&O\!\left(\!\frac{1}{(T\!\!+\!\!1)^{2u-2v-1}}\!+\!\frac{1}{(T\!\!+\!\!1)^{v\!-\!2\max\!\{w,0\}\!-\!1}}\!\!\right)\!.
\end{align}
Hence, substituting \eqref{eq71} and \eqref{eq72} into \eqref{eq70.5} implies\vspace{-0.6em}
\begin{align}\label{eq74}
 &\sum_{k=0}^{T}\E\|Y_k\|^2\cr
 =&O\!\left(\!\!(T\!\!+\!\!1)^{v}\!\!+\!\frac{1}{(T\!\!+\!\!1)^{2u\!-\!2v\!-\!1}}\!\!+\!\!\frac{1}{(T\!\!+\!\!1)^{v\!-\!2\max\!\{w\!,0\}\!-\!1}}\!\!\right)\!\!.
\end{align}{\vskip -3pt}

{\bf Step 6:} Finally, we prove $\liminf_{T\!\to\!\infty}\!\E\|\!\nabla\!F(x_{i,T\!+\!1})\!\|^2\!\!=\!\!0$ for any $i\!\in\!\mathcal{V}$. From Step 3, since there exists $G_1\geq0$ such that $\E(F(\bar{x}_T)-F^*)\leq G_1$ for any $T=0,1,\dots$, by Lemma \ref{lemma a1}(ii) we have
\begin{align}\label{eq75}
	\E\|\nabla F(\bar{x}_T)\|^2\leq 2L_1\E(F(\bar{x}_T)-F^*)\leq 2L_1G_1.
\end{align}{\vskip -4pt}\noindent Then, substituting \eqref{eq: optimization, 5} and \eqref{eq75} into \eqref{eq: optimization, 4} implies\vspace{-0.6em}
\begin{align}\label{eq76}
	&\E(F(\bar{x}_{k+1})-F^*)\cr
	\le&\E(F(\bar{x}_k)-F^*)-\frac{\alpha_T}{2}\E\|\nabla F(\bar{x}_k)\|^2\cr
	&+\frac{\alpha_TL_1^2(1+2\alpha_TL_1)}{2n}\E\|Y_k\|^2+\frac{\beta_T^2nrL_1}{2}\left(\Delta^2+\sigma_k^2\right)\cr
	&+\frac{\alpha_T^2\sigma_{\ell}^2L_1}{2\gamma_T}+2\alpha_T^2L_1^2G_1.
\end{align}
\noindent Note that \eqref{eq76} can be rewritten as\vspace{-0.4em}
\begin{align}\label{eq77}
	&\frac{\alpha_T}{2}\E\|\nabla F(\bar{x}_k)\|^2\cr
	\noalign{\vskip -4pt}
	\leq&\E(F(\bar{x}_k)-F(\bar{x}_{k+1}))+\frac{\alpha_TL_1^2(1+2\alpha_TL_1)}{2n}\E\|Y_k\|^2\cr
	&+\frac{\beta_T^2nrL_1}{2}\left(\Delta^2+\sigma_k^2\right)+\frac{\alpha_T^2\sigma_{\ell}^2L_1}{2\gamma_T}+2\alpha_T^2L_1^2G_1.
\end{align}
{\vskip -5pt}\noindent Then, since $F^*\leq F(x)$ holds for any $x\in\mathbb{R}^r$, summing \eqref{eq77} from $k=0$ to $T$ gives\vspace{-0.4em}
\begin{align*}
	&\frac{\alpha_T}{2}\sum_{k=0}^{T}\E\|\nabla F(\bar{x}_k)\|^2\cr
	\noalign{\vskip -9pt}
	\leq&\E(F(\bar{x}_0)-F(\bar{x}_{T+1}))+\frac{\alpha_TL_1^2(1+2\alpha_TL_1)}{2n}\sum_{k=0}^{T}\E\|Y_k\|^2
\end{align*}
\begin{align}\label{eq78}
	&+\sum_{k=0}^{T}\left(\frac{\beta_T^2nrL_1}{2}\left(\Delta^2\!+\!\sigma_k^2\right)\!+\!\frac{\alpha_T^2\sigma_{\ell}^2L_1}{2\gamma_T}\!+\!2\alpha_T^2L_1^2G_1\right)\cr
	\noalign{\vskip -7pt}
	\hspace{-1em}\leq&\E(F(\bar{x}_0)-F^*)+\frac{\alpha_TL_1^2(1+2\alpha_TL_1)}{2n}\sum_{k=0}^{T}\E\|Y_k\|^2\cr
	\noalign{\vskip -7pt}
	&+\sum_{k=0}^{T}\left(\frac{\beta_T^2nrL_1}{2}\left(\Delta^2\!+\!\sigma_k^2\right)\!+\!\frac{\alpha_T^2\sigma_{\ell}^2L_1}{2\gamma_T}\!+\!2\alpha_T^2L_1^2G_1\right).
\end{align}
{\vskip -7pt}\noindent By $\frac{1}{2}+\max\{w,0\}<v$ and $2u-v>1$ in Assumption \ref{assumption 4: step sizes}, \mbox{we have}\vspace{-0.6em}
\begin{align}\label{eq79}
	&\sum_{k=0}^{T}\left(\frac{\beta_T^2nrL_1}{2}\left(\Delta^2\!+\!\sigma_k^2\right)\!+\!\frac{\alpha_T^2\sigma_{\ell}^2L_1}{2\gamma_T}\!+\!2\alpha_T^2L_1^2G_1\right)\cr
	=&O\left(\sum_{k=0}^{T} \left(\frac{1}{(T\!\!+\!\!1)^{2v-2\max\{w,0\}}}+\frac{1}{(T\!\!+\!\!1)^{2u}}\right)\right)\cr
	=&O\left(\frac{1}{(T\!\!+\!\!1)^{2v-2\max\{w,0\}-1}}+\frac{1}{(T\!\!+\!\!1)^{2u-1}}\right).
\end{align}

{\vskip -7pt}\indent Note that $2u-v>1$ and $\frac{1}{2}+\max\{w,0\}<v<u$ in Assumption \ref{assumption 4: step sizes}. Then, we have $3u-2v-1=(2u-v-1)+(u-v)>0$, $u+v-2\max\{w,0\}-1>2v-2\max\{w,0\}-1>0$. For any $T$$=$$0$$,$$1$$,$$\dots$, substituting \eqref{eq74} and \eqref{eq79} into \eqref{eq78} implies\vspace{-0.6em}
\begin{align}\label{eq80}
	&\alpha_T\sum_{k=0}^{T}\E\|\nabla F(\bar{x}_k)\|^2\cr
	\noalign{\vskip -1pt}
	=&O\!\left(\!\frac{1}{(T\!\!+\!\!1)^{u-v}}\!+\!\frac{1}{(T\!\!+\!\!1)^{3u-2v-1}}\!+\!\frac{1}{(T\!\!+\!\!1)^{u+v-2\max\{w,0\}-1}}\!\right)\cr
	\noalign{\vskip -2pt}
	&+O\left(\frac{1}{(T\!\!+\!\!1)^{2v-2\max\{w,0\}-1}}+\frac{1}{(T\!\!+\!\!1)^{2u-1}}\right)\cr
	&+2(F(\bar{x}_0)-F(x^*)).
\end{align}{\vskip -8pt}\noindent Thus, there exists $G_2>0$ such that $\alpha_T\sum_{k=0}^{T}\E\|\nabla F(\bar{x}_T)\|^2\leq G_2$  for any $T=0,1,\dots$.

Next, we prove $\liminf_{T\to\infty}\E\|\nabla F(\bar{x}_{T+1})\|^2=0$ by contradiction. Suppose there exists $G_3>0$ such that $\liminf_{T\to\infty}\E\|\nabla F(\bar{x}_{T+1})\|^2=G_3>0$. Then, there exists a positive integer $T_1$ such that $\E\|\nabla F(\bar{x}_{T})\|^2\geq \frac{G_3}{2}$ for any $T=T_1,T_1+1,\dots$. Thus, we have\vspace{-0.7em}
\begin{align}\label{eq81}
	&\alpha_T\sum_{k=0}^{T}\E\|\nabla F(\bar{x}_{k})\|^2\geq\alpha_T\sum_{k=T_1}^{T}\E\|\nabla F(\bar{x}_{k})\|^2\cr
	\noalign{\vskip -3pt}
	\geq&\frac{\alpha_T(T\!-\!T_1\!+\!1)G_3}{2}\!\!=\!\!O\!\left(\!(T\!\!+\!\!1)^{1\!-\!u}\right)\!.
\end{align}
{\vskip -8pt}\noindent  Note that when $T$ goes to infinity, $\alpha_T\sum_{k=0}^{T}\E\|\nabla F(\bar{x}_{k})\|^2$ goes to infinity since the right hand side of \eqref{eq81} goes to infinity, which contradicts \eqref{eq80}. Then, we have $\liminf_{T\to\infty}\E\|\nabla F(\bar{x}_{T+1})\|^2=0$. Moreover, for any $i\in\mathcal{V}$, we have\vspace{-0.6em}
\begin{align}\label{eq82}
	&\E\|\nabla F(x_{i,T+1})\|^2\cr
	=&\E\|\nabla F(x_{i,T+1})-\nabla F(\bar{x}_{T+1})+\nabla F(\bar{x}_{T+1})\|^2\cr
	\leq&2\E\|\nabla\!F(x_{i,T\!+\!1})\!-\!\nabla\!F(\bar{x}_{T\!+\!1})\|^2\!+\!2\E\|\nabla\!F(\bar{x}_{T\!+\!1})\|^2\cr
	\leq&2L_1^2\E\|x_{i,T+1}-\bar{x}_{T+1}\|^2+2\E\|\nabla F(\bar{x}_{T+1})\|^2\cr
	\leq&2L_1^2\E\|Y_{T+1}\|^2+2\E\|\nabla F(\bar{x}_{T+1})\|^2.
\end{align}
{\vskip -6pt}\noindent Therefore, by $\lim_{T\to\infty}\E\|Y_{T+1}\|^2\hspace{-2pt}=\hspace{-2pt}0$ in Step 4, $\liminf_{T\to\infty}$ $\E\|\nabla F(x_{i,T+1})\|^2=0$ holds for any $i\in\mathcal{V}$. \hfill$\blacksquare$
\section{ Proof of Theorem \ref{thm3}}\label{appendix2}
\indent If Assumption \ref{asm5} holds, then \eqref{eq: optimization, 6} can be rewritten as\vspace{-0.6em}
\begin{align}\label{eq83}
	&\E\left(F(\bar{x}_{k+1})-F^*\right)
	\leq\left(1\!-\!\mu\alpha_T\!+\!2\alpha_T^2L_1^2 \right)\!\E(F(\bar{x}_k)\!-\!F^*)\nonumber\\
	&+\frac{\alpha_TL_1^2(1+2\alpha_TL_1)}{2n}\E\|Y_k\|^2+\frac{\alpha_T^2\sigma_{\ell}^2L_1}{2\gamma_T}\cr
	&+\frac{\beta_T^2 nrL_1}{2}(\Delta^2\!+\!\sigma_k^2)\!+\!2\alpha_T^2L_1^2M^*.
\end{align}
\indent For any $i\in\mathcal{V}$, by Lemma \ref{lemma a1}(i), we have
\begin{align}\label{eq: estimation,1}
	F(x_{i,T+1})\!-\!F(\bar{x}_{T+1})\le&\langle\nabla F(\bar{x}_{T+1}),x_{i,T+1}\!-\!\bar{x}_{T+1}\rangle\cr
	&+\frac{L_1}{2}\|\bar{x}_{T+1}-x_{i,T+1}\|^2.
\end{align}
Note that $\langle \mathbf{a},\mathbf{b}\rangle\le\|\mathbf{a}\|\|\mathbf{b}\|\le\frac{\|\mathbf{a}\|^2+\|\mathbf{b}\|^2}{2}$ for any $\mathbf{a},\mathbf{b}\in\mathbb{R}^{r}$. Then, \eqref{eq: estimation,1} can be rewritten as
\begin{align}\label{eq: estimation,2}
	&F(x_{i,T+1})-F(\bar{x}_{T+1})\cr
	\le& \frac{\|\!\nabla\! F(\bar{x}_{T\!+\!1})\|^2\!+\!\|\bar{x}_{T\!+\!1}\!-\!x_{i,T\!+\!1}\|^2}{2}\!+\!\frac{L_1}{2}\|\bar{x}_{T\!+\!1}\!-\!x_{i,T\!+\!1}\|^2\cr
	=&\frac{L_1+1}{2}\|\bar{x}_{T+1}-x_{i,T+1}\|^2+\frac{\|\nabla F(\bar{x}_{T+1})\|^2}{2}.
\end{align}
By Lemma \ref{lemma a1}(ii) we have $\|\nabla F(\bar{x}_{T\!+\!1})\|^2\!\leq\!2L_1(F(\bar{x}_{T\!+\!1})\!-\!F^*)$. This together with \eqref{eq: estimation,2} gives $F(x_{i,T\!+\!1})\!-\!F(\bar{x}_{T\!+\!1})\!\leq\!\frac{L_1\!+\!1}{2}\|\bar{x}_{T\!+\!1}\!-\!x_{i,T\!+\!1}\|^2\!+\!L_1(F(\bar{x}_{T\!+\!1})\!-\!F^*)$. Thus, we have
\begin{align}\label{eq: estimation,4}
	&F(x_{i,T+1})-F(\bar{x}_{T+1})\cr
	\le&\frac{L_1+1}{2}\sum_{i=1}^{n}\|\bar{x}_{T+1}-x_{i,T+1}\|^2+L_1(F(\bar{x}_{T+1})-F^*)\cr
	=&\frac{L_1+1}{2}\|Y_{T+1}\|^2+L_1(F(\bar{x}_{T+1})-F^*).
\end{align}

Furthermore, for any $i\in\mathcal{V}$, by \eqref{eq: estimation,4}, we have
\begin{align}\label{eq: estimation,5}
	&F(x_{i,T+1})-F^*\cr
	=&\left(F(x_{i,T+1})-F(\bar{x}_{T+1})\right)+\left(F(\bar{x}_{T+1})-F^*\right)\cr
	\le&\frac{L_1+1}{2}\|Y_{T+1}\|^2+(L_1+1)(F(\bar{x}_{T+1})-F^*)\cr
	\le&(L_1+1)\left(\|Y_{T+1}\|^2+(F(\bar{x}_{T+1})-F^*)\right).
\end{align}
Let \vspace{-0.4em}
\begin{align}\label{eq89}
	\theta_5=&\max\{1\!-\!\mu\alpha_T+2\alpha_T^2L_1^2+\frac{4n(1+\rho_{\mathcal{L}}\beta_T)\alpha_T^2L_1}{\rho_{\mathcal{L}}\beta_T},\cr
	&~~1\!-\!\rho_{\mathcal{L}}\beta_T\!\!+\!\!\frac{\alpha_TL_1^2\!(1\!+\!2\alpha_TL_1)}{2n}\!+\!\frac{2(1\!\!+\!\!\rho_{\mathcal{L}}\beta_T)\alpha_T^2\!L_1^2}{\rho_{\mathcal{L}}\beta_T}\!\}\!.
\end{align}
{\vskip -2pt}\noindent Then, by \eqref{eq: theta 2} and \eqref{eq89}, summing  \eqref{eq:consensus ,4}  and  \eqref{eq83} implies\vspace{-0.4em}
\begin{align}\label{eq90}
	&\E(\|Y_{k+1}\|^2+F(\bar{x}_{k+1})-F^*)\cr
	\leq&\theta_5\E(\|Y_k\|^2+F(\bar{x}_k)-F^*)+\theta_{k,2}.
\end{align}
{\vskip -4pt}\noindent Thus, by \eqref{eq: estimation,5}, iteratively computing \eqref{eq90} gives\vspace{-0.4em}
\begin{align}\label{eq: inequality, 1}
	&\E\!(F(x_{i,T\!+\!1})\!-\!F^*\!)\!\cr
	\leq&(L_1+1)\E(\|Y_{T+1}\|^2+F(x_{T\!+\!1})\!\!-\!\!F^*)\cr
	\noalign{\vskip -3pt}
	\leq&\theta_5^{T\!+\!1}\!(L_1\!\!+\!\!1)(\|Y_0\|^2\!\!+\!\!F(\bar{x}_0)\!\!-\!\!F^*)\!\!+\!\!(L_1\!\!+\!\!1)\!\sum_{k=0}^{T}\!\!\theta_5^{T\!-\!k}\theta_{k,2}.
\end{align}
By Assumption \ref{assumption 4: step sizes}, we have $\theta_5>0$.  Then, we have $\theta_5^{T\!+\!1}\!=\!\exp((T\!+\!1)\ln (1\!-\!(1\!-\!\theta_5)))\!=\!O(\exp(-(T\!+\!1)(1\!-\!\theta_5)))$.  This together with \eqref{eq89} implies
\begin{align}\label{eq: inequality, 2}
	&\theta_5^{T+1}\cr
	=&O\!\!\left(\!\!\max\{\!\exp(-\!(T\!\!+\!\!1)\!\mu\alpha_T\!\!+\!\!(T\!\!+\!\!1)\!(2\alpha_T^2\!L_1^2\!\!+\!\!\frac{4n\!(1\!\!+\!\!\rho_{\mathcal{L}}\beta_T)\alpha_T^2L_1}{\rho_{\mathcal{L}}\beta_T})\!),\right.\cr
	&\exp(-(T\!\!+\!\!1)\rho_{\mathcal{L}}\beta_T\cr
	&\left.+(T\!\!+\!\!1)\!(\frac{\alpha_TL_1^2(1\!+\!2\alpha_TL_1)}{2n}\!+\!\frac{2(1\!+\!\rho_{\mathcal{L}}\beta_T)\alpha_T^2L_1^2}{\rho_{\mathcal{L}}\beta_T})\!)\!\}\!\right).
\end{align}
Note that $u>v>\frac{1}{2}+\max\{w,0\}$ in Assumption \ref{assumption 4: step sizes}. Then, we have $-\rho_{\mathcal{L}}\beta_T\!+\!\frac{\alpha_TL_1^2\!(1\!+\!2\alpha_TL_1)}{2n}\!+\!\frac{2(1+\rho_{\mathcal{L}}\beta_T)\alpha_T^2L_1^2}{\rho_{\!\mathcal{L}}\beta_T}\!=\!O(-\frac{\rho_{\mathcal{L}}}{2}\beta_T)$, and $-\mu\alpha_T\!+\!2\alpha_T^2L_1^2\!+\!\frac{4n(1+\rho_{\mathcal{L}}\beta_T)\alpha_T^2L_1}{\rho_{\mathcal{L}}\beta_T}\!=\!O(-\frac{\mu}{2}\alpha_T)$. Thus, by $2u-v>1$ in Assumption~\ref{assumption 4: step sizes}, \eqref{eq: inequality, 2} can be rewritten as
\begin{align}\label{eq: inequality, 5}
	\hspace{-1.2em}&\theta_5^{T+1}\cr
	\hspace{-1.2em}=&O\!\left(\!\max\{\exp(-(T\!\!+\!\!1)\frac{\mu}{2}\alpha_T),\exp(-(T\!\!+\!\!1)\frac{\rho_{\mathcal{L}}}{2}\beta_T)\}\right)\cr
	\hspace{-1.2em}=&O\!\left(\!\max\{\exp\!\left(\!-\!\frac{\mu a_1}{2} (T\!\!+\!\!1)^{1\!-\!u}\!\right)\!\!,\exp\!\left(\!-\!\frac{\rho_{\mathcal{L}}a_2}{2}(T\!\!+\!\!1)^{1\!-\!v}\!\right)\!\}\!\right)\!.	
\end{align}
Similar to \eqref{eq62}-\eqref{eq64}, we have
\begin{align}\label{eq: inequality, 9}
	\sum_{k=0}^{T}\!\theta_5^{T\!-\!k}\theta_{k,2}\!=&O\!\left(\!\!\frac{\Delta^2}{(T\!\!+\!\!1)^{2v\!-\!2\max\{w,0\}\!-\!1}}\!\!+\!\!\frac{1}{(T\!\!+\!\!1)^{2u\!-\!v\!-\!1}}\!\!\right)\!\cr
	=&O\!\left(\!\!\frac{\Delta^2}{(T\!\!+\!\!1)^{\min\{2v-2\max\{w,0\}-1,2u-v-1\}}}\!\!\right)\!.
\end{align}
Hence, by substituting \eqref{eq: inequality, 5} and \eqref{eq: inequality, 9} into \eqref{eq: inequality, 1}, we have
\begin{align}\label{eq95}
	\hspace{-1.3em}\E\!(F(x_{i,T\!+\!1})\!-\!F^*\!)\!\!=\!\!O\!\!\left(\!\!\frac{\Delta^2}{(T\!\!+\!\!1)^{\min\{2v\!-\!2\max\{w,0\}\!-\!1,2u\!-\!v\!-\!1\}}}\!\!\right)\!.
\end{align}
\indent Note that by Lemma \ref{lemma a1}(ii), we have 
\begin{align}\label{eq96}
	\|\nabla F(x_{i,T+1})\|^2\leq 2L_1(F(x_{i,T+1})-\!F^*).
\end{align}
Then, taking the mathematical expectation on \eqref{eq96} and substituting \eqref{eq95} into \eqref{eq96} imply
\begin{align}\label{eq97}
	\hspace{-2em}\E\|\nabla F(x_{i,T\!+\!1})\|^2\leq& 2L_1\E(F(x_{i,T\!+\!1})\!-\!F^*)\cr
	=&O\!\!\left(\!\!\frac{\Delta^2}{(T\!\!+\!\!1)^{\min\{2v\!-\!2\max\{w,0\}\!-\!1,2u\!-\!v\!-\!1\}}}\!\!\right)\!\!.
\end{align}
\indent Note that for any $\psi\in[1,2]$, the function $x^{\frac{\psi}{2}}$ is concave in $x$. Then, by Jensen's inequality \cite[Cor. 4.3.1]{chow2012probability} we have $\E\!\|\!\nabla F(x_{i,T\!+\!1})\|^\psi=\E(\|\nabla F(x_{i,T+1})\|^2)^{\frac{\psi}{2}}\leq(\E\|\nabla F(x_{i,T\!+\!1})\|^2)^{\frac{\psi}{2}}$. This together with \eqref{eq97} implies $\E\|\!\nabla F(x_{i,T\!+\!1})\|^\psi$$=$$O(\!\frac{\Delta^2}{(T\!+\!1)^{\frac{\psi}{2}\min\{2v\!-\!2\max\{w,0\}\!-\!1,2u\!-\!v\!-\!1\}}}\!)$, where the constant in the big-$O$ notation does not depend on $\Delta$. \hfill$\blacksquare$

\end{document}